 \documentstyle[aps,prc,twocolumn,floats,epsfig]{revtex}
 \draft
\begin{document}

\renewcommand{\thefootnote}{\arabic{footnote}}

\twocolumn[\columnwidth\textwidth\csname@twocolumnfalse\endcsname

\title{Shell structure of superheavy nuclei 
       in self-consistent mean-field models} 

\author{M. Bender${}^{1,2,3}$,
        K. Rutz${}^{1}$,  
        P.-G. Reinhard${}^{4,5}$, 
        J. A. Maruhn${}^{1,5}$,
        and W. Greiner${}^{1,5}$
}

\address{${}^{1}$Institut f\"ur Theoretische Physik,
         Universit\"at Frankfurt,
         Robert-Mayer-Strasse 10, D-60325 Frankfurt, Germany
}

\address{${}^{2}$Department of Physics and Astronomy,
         University of North Carolina, 
         Chapel Hill, NC 27599-3255
}

\address{${}^{3}$Department of Physics and Astronomy, 
         University of Tennessee,
         Knoxville, TN 37996
}

\address{${}^{4}$Institut f\"ur Theoretische Physik II,
         Universit\"at Erlangen-N\"urnberg,
         Staudtstrasse 7, D-91058 Erlangen, Germany
}

\address{${}^{5}$Joint Institute for Heavy--Ion Research, 
         Oak Ridge National Laboratory,
         P. O. Box 2008, Oak Ridge, Tennessee 37831
}

\date{March 25 1999}

\maketitle

\addvspace{5mm}

%
%
%
\begin{abstract}
We study the extrapolation of nuclear shell structure to the region 
of superheavy nuclei in self-consistent mean-field models -- the 
Skyrme-Hartree-Fock approach and the relativistic mean-field model -- 
using a large number of parameterizations which give similar results
for stable nuclei but differ in detail. Results obtained with the 
Folded-Yukawa potential which is widely used in macroscopic-macroscopic 
models are shown for comparison.
We focus on differences in the isospin dependence of the spin-orbit 
interaction and the effective mass between the models and their 
influence on single-particle spectra. 
The predictive power of the mean-field models concerning single-particle 
spectra is discussed for the examples of ${}^{208}{\rm Pb}$ and the 
spin-orbit splittings of selected neutron and proton levels in 
${}^{16}{\rm O}$, ${}^{132}{\rm Sn}$ and ${}^{208}{\rm Pb}$.
While all relativistic models give a reasonable description of 
spin-orbit splittings, all Skyrme interactions show a wrong trend
with mass number. The spin-orbit splitting of heavy nuclei might be 
overestimated by \mbox{$40\%-80\%$}, which exposes a fundamental deficiency 
of the current non-relativistic models.
In most cases the occurrence of spherical shell closures is found to be
nucleon-number dependent. 
Spherical doubly-magic superheavy nuclei are found at 
\mbox{${}^{298}_{184}114$}, \mbox{${}^{292}_{172}120$}, 
or \mbox{${}^{310}_{184}126$} depending on the parameterization.
The \mbox{$Z=114$} proton shell closure, which is related to 
a large spin-orbit splitting of proton $2f$ states, is predicted only
by forces which by far overestimate the proton spin-orbit splitting in
${}^{208}{\rm Pb}$.
The \mbox{$Z=120$} and \mbox{$N=172$} shell closures predicted by the 
relativistic models and some Skyrme interactions are found to be related 
to a central depression of the nuclear density distribution. 
This effect cannot appear in macroscopic-microscopic models or 
semi-classical approaches like the extended Thomas--Fermi-Strutinski 
integral approach which have a limited freedom for the density distribution 
only.
In summary, our findings give a strong argument for \mbox{${}^{292}_{172}120$}
to be the next spherical doubly-magic superheavy nucleus.
\end{abstract}


\addvspace{5mm}]

\narrowtext
%
%
\section{Introduction}
The extrapolation of nuclear shell structure to superheavy systems
has been discussed since the early days of the shell correction method
\cite{Mye66a,SuperNils,Mosel,Fiz72a,Bra72a}, when spherical proton shell 
closures at \mbox{$Z = 114$} and \mbox{$Z = 126$} and a spherical 
neutron shell closure at \mbox{$N = 184$} were predicted. Shell
effects are crucial for the stability of superheavy nuclei which by
definition have a negligible liquid-drop fission barrier.
Recent experimental progress allowed the synthesis of three new 
superheavy elements with \mbox{$Z=110$}--$112$
\cite{Z111,Berkley,Z112,Dubna2,SHReview}, but these nuclides are believed 
to be well deformed. The experimental data on these nuclei and their
decay products -- $\alpha$-decay half-lives and $Q_\alpha$ values -- 
agree with the theoretical prediction 
\cite{Sob87,Sob89,Pat89a,Patyk,Mol92a,Mol94a} 
of a deformed neutron shell at \mbox{$N = 162$} which has a significant 
stabilizing effect \cite{SHReview,Laz94}. The experimental proof of 
the deformed
shell by a measurement of the deformation is beyond the current 
experimental possibilities. As a first step in this direction the 
ground-state deformation of ${}^{254}{\rm No}_{102}$ was deduced from 
its ground-state rotational band in a recent experiment \cite{Reiter}.
The ultimate goal is to reach the expected island of spherical 
doubly-magic superheavy nuclei. More refined parameterizations of 
macroscopic-microscopic models \cite{Pat89a,Patyk,Mol92a,Mol94a} 
confirm the older finding that it is located around ${}^{298}_{184}114$. 
These nuclei, although even heavier than the heaviest nuclides known so far,
are expected to have much longer half-lives due to the stabilizing effect 
of the spherical shell closure which significantly increases the 
fission barriers \cite{Cwi83,Cwi85,Bon86a,Smo95a}.

Although modern macroscopic-microscopic models quite successfully 
describe the bulk properties of known nuclei throughout the chart 
of nuclei, their parameterization needs preconceived knowledge about 
the density distribution and the nuclear potentials which fades away 
when going to the limits of stability.
Like the mean-field models based on the shell correction method,
self-consistent mean-field models have been used for the 
investigation of superheavy nuclei from the earliest parameterizations 
\cite{Vau70b,Bei74a} to the most recent ones
\cite{Ton80,Gam90,Boe93,Berger,Ring,Naz,RutzSuper,BuervenSuper,SHpes}.

In two previous articles we have discussed the occurrence of spherical 
\cite{RutzSuper} and deformed \cite{BuervenSuper} shell closures in 
superheavy nuclei for a large number of parameterizations of self-consistent 
nuclear structure models, namely the Skyrme-Hartree-Fock (SHF) approach 
\cite{refSHF}, and the relativistic mean-field (RMF) model 
\cite{WalSer,Rei89,Rin95}.  
Spherical proton shell closures are predicted for \mbox{$Z = 114$}, 
\mbox{$Z = 120$} and \mbox{$Z = 126$}, depending on the parameterization, 
while neutron shell closures occur at \mbox{$N = 172$} and \mbox{$N = 184$} 
respectively. Only one parameterization -- the Skyrme interaction SkI4 -- 
confirms the prediction of macroscopic-microscopic models for a 
doubly magic ${}^{298}_{184}114$, other parameterizations -- the Skyrme
forces SkM* and SkP -- predict ${}^{310}_{184}126$, while yet others 
-- the Skyrme interaction SkI3 and most of the 
relativistic forces -- give a new alternative with ${}^{292}_{172}120$. 
Several interactions predict no doubly magic spherical superheavy nucleus 
at all. In self-consistent models, the proton and neutron shells strongly
affect each other \cite{RutzSuper}. 
Small details of the shell structure have a strong 
influence on the potential energy surfaces of superheavy nuclei in the 
vicinity of the ground-state deformation, leading to dramatic differences 
in the fission barrier heights and therefore in the fission half-lives,
while the predictions of different models and forces are similar at 
large deformations \cite{SHpes}.

Superheavy nuclei differ from stable nuclei by their larger charge and 
mass numbers. The strong Coulomb potential induces significant changes 
in the proton shell structure: single-particle states with large
angular momentum and small overlap with the nuclear center only are 
lowered compared to small-$j$ states, see Figs.~1--2 of Ref.~\cite{Naz} 
and the discussion therein. While this effect occurs already in 
non-self-consistent models, polarization effects of the density 
distribution due to the high charge number can be described in 
self-consistent models only. The Coulomb interaction pushes protons 
to larger radii, which changes the density distribution and 
the single-particle potentials of both protons and neutrons in a
complicated manner. On the other hand, the large mass number of superheavy
nuclei leads to a high average density of single-particle levels. 
Therefore the search for shell effects in superheavy nuclei probes the 
detailed relations among the single-particle states with extremely high
sensitivity. 

The question arises which features of the effective 
mean-field models are most decisive for the single-particle structure. 
The three most crucial ingredients in this respect are: 
first the effective nucleon mass and its radial dependence
which determines the level density near the Fermi surface, 
second the spin-orbit potential
which determines the energetic distance of the spin-orbit partners,
and third the density dependence of potential and effective mass which
has an influence on the relative position of the states. 
We perform here a comparison of various parameterizations from SHF 
as well as RMF with emphasis on their spin-orbit properties. 
The effective masses (with one exception) are comparable in all 
forces. The density dependences are
similar amongst the SHF forces and amongst the RMF forces, but
differ significantly between SHF and RMF. The largest variations
in the sample of parameterizations occurs indeed for the spin-orbit 
part of the forces where we have three classes, the standard SHF models,
SHF with extended spin-orbit forces (SkI3 and SkI4), and the
RMF models. The present paper concentrates predominantly on this
given variation of the spin-orbit force. 
It is the aim of this paper to explain the contradicting results
of self-consistent models mentioned above and to find the most
reliable prediction for the next spherical doubly-magic superheavy
nucleus.

In Sect.~\ref{Sect:frame} the properties of the mean-field models and
the parameterizations used is discussed. In Sect.~\ref{Sect:LS} the
details of the spin-orbit interaction and the differences between 
the various models used are explained, while Sect.~\ref{Sect:effmass} 
discusses briefly the relation between effective mass and average 
density of single-particle levels.
In Sect.~\ref{Sect:magic} we compare the predictions of the 
various mean-field models with known single-particle energies
in \mbox{${}^{208}{\rm Pb}$} and experimental spin-orbit splittings
in \mbox{${}^{16}{\rm O}$}, \mbox{${}^{132}{\rm Sn}$} and 
\mbox{${}^{208}{\rm Pb}$} and study the shell structure of the potential
spherical doubly-magic nuclei \mbox{${}^{298}_{184}114$}, 
\mbox{${}^{292}_{172}120$}, and \mbox{${}^{310}_{184}126$}
and the predicted nucleon-number dependence of the 
\mbox{$Z=120$} proton shell and the \mbox{$N=172$} neutron shell 
in some detail. Sect.~\ref{Sect:summary} summarizes our findings.
In an appendix we present the details of the 
mean-field and pairing models necessary for our discussion.
%
%
\section{The framework}
\label{Sect:frame}
The Skyrme force was originally designed as an effective two-body 
interaction for self-consistent nuclear structure calculations. It has the 
technical advantage that the exchange terms in the Hartree-Fock equations 
have the same form as the direct terms and therefore the
numerical solution of the Skyrme-Hartree-Fock equations is as simple as
in case of the Hartree approach, while the solution of the
Hartree-Fock equations using finite-range forces like the Gogny
force \cite{Gogny} is a numerically challenging task. The total binding 
energy can be formulated in terms of an energy functional which depends 
on local densities and currents only, see Appendix~\ref{Sect:Parameters}. 
This links the Skyrme-Hartree-Fock model to the effective energy functional 
theory in the Kohn-Sham approach which was originally developed for
many-electron systems. The Hohenberg-Kohn theorem \cite{HohKohn} states 
that the non-degenerate ground-state energy of a many-Fermion system with 
local two-body interactions is a unique functional of the local density 
only. The Kohn-Sham scheme \cite{KohnSham} relies on the Hohenberg-Kohn 
theorem but keeps the full dependence on the single-particle wavefunctions 
for the kinetic energy which allows to preserve the full shell
structure while employing for the rest rather simple functionals in
local-density approximation. This point of view can be carried over
to the case of nuclei where, however, the non-local two-body
interaction requires an extension of the energy functional by a dependence 
on other densities and currents, e.g., the spin-orbit current. In any
case, there is no need for a fundamental two-body force in an effective 
many-body theory, but one can start from an effective energy functional 
which is formulated directly at the level of one-body densities and 
currents (see, e.g., \cite{ReiCorr} and references therein).

The relativistic mean-field model can be seen from the same point of view 
as a relativistic generalization of the non-relativistic models
using a finite-range interaction formulated in terms of effective 
mesonic fields. Relativistic kinematics plays no role
in nuclear structure physics, but the RMF naturally describes the
spin-orbit interaction in nuclei, which is a relativistic effect
that has to be added phenomenologically in non-relativistic models. 
This will be discussed in Sect.~\ref{Sect:LS} in more detail.

For both SHF and RMF there are numerous parameterizations in the
literature. We select here a few typical samples of comparable (high)
quality, mostly from recent fits.  For the non-relativistic
Skyrme-Hartree-Fock calculations we consider the Skyrme forces SkM*
\cite{SkM*}, SkP \cite{SkP}, SLy6, SLy7 \cite{Chabanat,SLyx}, SkI1, SkI3, and
SkI4 \cite{SkIx}. For the RMF we consider NL3 \cite{NL3}, NL-Z,
\cite{NLZ}, and two completely new forces, NL-Z2 and NL-VT1.  
All forces are developed through fits to given nuclear data, but with
different bias. Of course, the basic ground-state properties of
spherical nuclei (energy, radius) are always well reproduced. Small
variations appear with respect to further demands.
The parameterization ${\rm SkM}^\ast$ is the oldest in the list here.
It was the first Skyrme force with acceptable incompressibility as well 
as fission properties and remains up to date a reliable parameterization 
in several respects.
The Skyrme force SkP was developed around the same time with the aim
to allow the simultaneous description of mean-field and pairing
channel. Moreover, it was decided here to use effective mass
\mbox{$m^\ast/m = 1.0$}.  (Mind that all other forces in our sample have
smaller effective masses around \mbox{$0.6 \leq m^\ast/m \leq 0.8$}).
The forces SLy6 and SLy7 stem from a series of fits where it was
successfully attempted to cover properties of neutron matter together
with normal nuclear ground-state properties. In SLy6 the contribution
of the kinetic terms of the Skyrme force to the spin-orbit potential
is discarded, which is common practice for nearly all Skyrme
parameterizations, e.g.\ SkM* and the SkI$x$ forces in the sample
here. SLy7 is fitted exactly in the same way as SLy6, but these
additional contributions to the spin-orbit force are considered, see
the discussion in Sect.~\ref{Subsect:LSfield} for details.
The forces SkI1, SkI3 and SkI4 stem from a recent series of fits along
the strategy of \cite{SkyrmeFit} where additionally key features of
the nuclear charge formfactor were included providing information on
the nuclear surface thickness. For these, furthermore, information from 
exotic nuclei was taken into account in order to better determine the
isotopic parameters. The force SkI1 is a fit within the standard
parameterization of the Skyrme forces. This performs very well in all
respects, except for the isotopic trends of the charge radii in the
lead region. To cover these data, one needs to extend the spin-orbit
functional by complementing it with an additional isovector degree of
freedom \cite{SkIx} as will be discussed in
Sect.~\ref{Subsect:LSfield} in more detail. SkI3 uses a fixed
isovector part built in analogy to the RMF, whereas SkI4 was fitted
allowing free variation of the isovector spin-orbit force.  The
modified spin-orbit force has a strong effect on the spectral
distribution in heavy nuclei and thus even more influence for the
predictions of shell closures in the region of superheavy nuclei.

The forces headed by ``NL'' belong to the domain of the RMF
model.  The parameterizations NL-Z, NL-Z2, and NL3 use the standard
nonlinear \emph{ansatz} for the RMF model, whereas NL-VT1 additionally
considers a tensor coupling of the vector mesons. The parameterization
NL-Z \cite{Rei89} aims at a best fit to nuclear ground-state properties
along the strategy of \cite{SkyrmeFit}. It is a re-fit of
the popular force NL1 with a microscopic treatment of the correction
for spurious center-of-mass motion. NL-Z2 and NL-VT1 are new
parameterizations developed for the purpose of these studies to match
exactly the same enlarged set of data including information on exotic
nuclei like the SkI$x$ Skyrme forces. This should allow better
comparison between the RMF and the Skyrme model. The force NL3, finally,
results from a recent fit including neutron rms radii. It gives a
good description of both nuclear ground states and giant
resonances. Details of the RMF Lagrangian and the actual
parameterizations are discussed in Appendix~\ref{Subsect:RMF}.
%
%
\mediumtext
\begin{table*}[t!]
\caption{
\label{Tab:NucMat}
Compilation of nuclear matter properties for the parameter sets used in 
this study.
$E/A$ and $\rho_{0}$ denote the equilibrium energy per nucleon 
and density, $K_{\infty}$ the compression modulus, $m^\ast/m$ the 
effective mass in units of the free mass (note that we provide
two values for the relativistic models where the value in brackets 
is $m^\ast/m(k_F)$ at the Fermi surface and the other at $k\!=\!0$
\protect\cite{JamMah})
$a_{\rm sym}$ the asymmetry coefficient, and
$\kappa$ the sum--rule enhancement factor.}
\begin{tabular}{lcddcdc}
Force      & $E/A \, [{\rm MeV}]$
           & $\rho_0 \, [{\rm fm}^{-3}]$
           & $K_\infty \,[{\rm MeV}]$
           & $m^\ast/m$
           & $a_{\rm sym}$ 
           & $\kappa$ \\ \tableline
SkP        & $-16.04$ & 0.163 & 202 & 1.000 & 30.0 & 0.35  \\
SkM$^\ast$ & $-16.01$ & 0.160 & 217 & 0.789 & 30.0 & 0.53  \\
SLy6       & $-15.92$ & 0.159 & 230 & 0.690 & 32.0 & 0.25  \\
SLy7       & $-15.90$ & 0.158 & 230 & 0.688 & 32.0 & 0.25  \\
SkI1       & $-15.93$ & 0.160 & 243 & 0.693 & 37.5 & 0.25  \\
SkI3       & $-15.96$ & 0.158 & 258 & 0.577 & 34.8 & 0.25  \\
SkI4       & $-15.92$ & 0.160 & 248 & 0.650 & 29.5 & 0.25  \\ \tableline
NL3        & $-16.24$ & 0.148 & 272 & 0.595 (0.659) & 37.4 & 0.68 (0.53) \\
NL--Z      & $-16.18$ & 0.151 & 173 & 0.583 (0.648) & 41.7 & 0.72 (0.55) \\
NL--Z2     & $-16.07$ & 0.151 & 172 & 0.583 (0.648) & 39.0 & 0.72 (0.55) \\
NL--VT1    & $-16.10$ & 0.150 & 179 & 0.600 (0.663) & 39.0 & 0.66 (0.51) \\
\end{tabular}
\end{table*}
\narrowtext
%
%

The nuclear matter properties of the forces are summarized in
Table~\ref{Tab:NucMat}. These are to be considered mainly as
extrapolations from finite nuclei to the infinite system.  There a few
exceptions because in some cases the one or the other nuclear matter
property has entered as a constraint into the fit. These cases are:
the effective mass \mbox{$m^\ast/m = 1$} for SkP, the compressibility
\mbox{$K_\infty = 230 \; {\rm MeV}$} and asymmetry coefficient
\mbox{$a_{\rm sym} = 32.0 \; {\rm MeV}$} for the SLy$x$ forces, 
and the sum-rule
enhancement factor \mbox{$\kappa = 0.25$} in case of the SLy$x$ and
SkI$x$ forces. Table~\ref{Tab:NucMat} shows that most Skyrme
forces share the basic nuclear matter properties close to the
phenomenological values like binding energy per nucleon \mbox{$E/A
\approx -16 \; {\rm MeV}$}, equilibrium density
\mbox{$\rho_0 \approx 0.16 \; {\rm fm}^{-3}$}, incompressibility 
$K=210\pm30\,{\rm MeV}$ \cite{Blaizot}, 
asymmetry energy \mbox{$30 \, {\rm MeV} \leq a_{\rm sym} \leq 32
\,{\rm MeV}$}, and a low sum-rule enhancement factor \mbox{$0 \leq
\kappa \leq 0.25$}. A phenomenological value for the
effective mass of $m^*/m\approx 0.8$
can be drawn from the position of the giant quadrupole
resonance in heavy nuclei \cite{BGHrev}. And we see that 
the mean field results for the effective mass vary in a wide range
\mbox{$0.58 \leq m^\ast/m \leq 1.0$} about this value.
This is a bit disquieting because the effective mass
is a feature which
has a strong impact on spectral properties, influencing, in turn, the
predictions for superheavy nuclei.

The nuclear matter properties of the relativistic parameterizations
differ significantly from those of Skyrme forces. $E/A$ is usually
slightly larger and $\rho_0$ somewhat smaller than the values for
Skyrme interactions. The predictions for the incompressibility $K$ differ
systematically from those 
of the nonrelativistic models, in case of NL3 it is somewhat
larger, in case of the other RMF forces smaller than the
average result for Skyrme forces. But all parametrizations stay within the
accepted bounds of this rather uncertain quantity.
The asymmetry coefficient and the sum-rule enhancement factor
are substantially larger than in case of the Skyrme forces.
But all RMF forces agree in their rather low value for the
effective mass \mbox{$0.58 \leq m^\ast/m \leq 0.6$}. It is to be
noted, however, that the effective mass in RMF depends on the momentum
as
\begin{eqnarray}
\label{eq:mstarRMF}
\frac{m^*(k_F)}{m}
& = & \sqrt{\left(\frac{m^*(0)}{m}\right)^2+\left(\frac{k_F}{m}\right)^2}
      \nonumber \\
& \approx & \sqrt{\left(\frac{m^*(0)}{m}\right)^2+0.08}
\end{eqnarray}
where $m^*(0)$ is the value at $k=0$ usually handled as effective mass
in the RMF  and where we assumed in the second step
a typical $k_F\approx 1.35/{\rm fm}$.
Table~\ref{Tab:NucMat} thus shows two values for $m^*/m$ 
in case of the RMF, at momentum zero and in
brackets the more relevant value $m^*/m(k_F)$ at the Fermi surface.
The latter value is larger by about $10\%$ 
and comes visibly closer to the results for the Skyrme forces.

In view of the application to superheavy nuclei, it is worthwhile to
check the performance of all these forces in our sample with respect
to already known superheavy nuclei. This was done in 
Ref.\ \cite{BuervenSuper}. It turns out, that SkI3, SkI4, and the 
relativistic forces perform best in that respect, although it is to be 
mentioned that all relativistic forces show a wrong isotopic trend, see 
\cite{BuervenSuper} for details. It is noteworthy that the extended 
Skyrme functionals SkI3 and SkI4 perform much better in the region of 
superheavy nuclei than the Skyrme parameterizations with the standard 
spin-orbit interaction. This indicates that an extended spin-orbit 
interaction is an essential ingredient for the description of heavy systems.

In both SHF and RMF the pairing correlations are treated in the BCS 
scheme using a delta pairing force, see Appendix~\ref{Subsect:PairFunc} 
for details.

The numerical procedure solves the coupled SHF and RMF equations 
on a grid in coordinate space with the damped gradient iteration 
method \cite{dampgrad}. The codes for the solution of both 
SHF and RMF models have been implemented in a common programming 
environment sharing all the crucial basic routines.
%
%
\section{Spin-orbit interaction in nuclear mean-field models}
\label{Sect:LS}
\subsection{The spin-orbit field}
\label{Subsect:LSfield}
The spin-orbit interaction is an essential ingredient of every model dealing 
with nuclear shell structure to explain the shell closures of heavy nuclei 
beyond \mbox{$N=Z=20$} \cite{LS1,LS2}. It was already noted in the first 
explorations with the modified oscillator model  
that different fits of the spin-orbit coupling constant lead 
to contradicting predictions for the next major shell closures in
superheavy nuclei \cite{Nilssonbook}.

The spin-orbit interaction emerges naturally in relativistic models and
the explanation of the large spin-orbit splitting in nuclei was one of the
first prominent successes of the relativistic mean-field approach
\cite{Duerr}. The spin-orbit potential can be deduced 
in the non-relativistic limit of the RMF and is given up to order 
$v^2/c^2$ by \cite{Rei89}
\begin{equation}
\label{eq:LS:rmf1}
{\bf W}_q^{({\rm rmf})}
\approx - \frac{\hbar c}{(2 m_q + {\cal S}_q - {\cal V}_q)^2} 
          \nabla ( {\cal S}_q - {\cal V}_q )
\quad ,
\end{equation}
where ${\cal S}$ and ${\cal V}$ are the scalar and vector potentials
respectively, see appendix~\ref{Subsect:RMF} for details.
While the usual potential is given by the sum
of the large negative scalar potential ${\cal S}$ and the large positive
vector potential ${\cal V}$ which cancel nearly to give the usual
shell-model potential, the difference of scalar and vector potential enters 
the expression for the spin-orbit field, explaining its large strength.
The occurrence of the derivative of the fields in (\ref{eq:LS:rmf1})
indicates that the spin-orbit field is peaked in the nuclear surface 
region and that its strength will depend on the surface thickness
of the particular nucleus.

To compare with the corresponding expression for Skyrme interactions, one 
has to evaluate (\ref{eq:LS:rmf1}) in local density approximation
\begin{equation}
\label{eq:W_rmflda}
{\bf W}_q^{{(\rm rmf})}
\approx \frac{\hbar c}{(2 m_q - C \rho - C' \rho_q)^2} \; 
        ( C \nabla \rho + C' \nabla \rho_q )
\quad ,
\end{equation}
where \mbox{$C = C_\sigma + C_\omega - C_\rho$} and 
\mbox{$C^\prime = 2 C_\rho$} are combinations of RMF parameters
with \mbox{$C_i = g_i^2/m_i^2$}. The isospin dependency of the spin-orbit 
potential is rather weak for typical RMF parameterizations which give 
\mbox{$C^\prime \approx 0.1 \times C$}.

In the framework of non-relativistic models the zero-range two-body 
spin-orbit interaction proposed  by Bell and Skyrme \cite{Bell,Sky59b} is 
widely used. Examples are all standard Skyrme interactions like SkM*, SkP,
the SLy$x$ forces or SkI1 and other non-relativistic effective interactions 
like the Gogny force \cite{Gogny}.
The corresponding spin-orbit potential ${\bf W}_q$ is given by
\begin{equation}
\label{eq:W_std}
{\bf W}_q^{{(\rm std})}
= b_4 ( \nabla \rho + \nabla \rho_q )
\quad .
\end{equation}
There are two fundamental differences between the relativistic and
non-relativistic expressions for the spin-orbit potential: the
isospin dependence and the missing density dependence in case of the
non-relativistic models.

When deriving the single-particle Hamiltonian from an underlying
Skyrme force there appears an additional contribution to the spin-orbit 
field which arises from the momentum-dependent terms 
in the two-body Skyrme force
\begin{equation}
\label{eq:W_J}
{\bf W}_q^{({\bf J})}
= b_4 ( \nabla \rho + \nabla \rho_q)
  + c_1  {\bf J}
  - c_1' {\bf J}_q  
\quad .
\end{equation}
The calculation of the spin-orbit current ${\bf J}$ is somewhat cumbersome
in deformed codes and its contribution to the total binding energy 
rather small. Therefore the ${\bf J}$-dependent terms in (\ref{eq:W_J})
are discarded in most parameterizations of the Skyrme interaction
and (\ref{eq:W_std}) is used instead. SkP and SLy7 are two exceptions 
in this investigation.

In the Hohenberg-Kohn-Sham interpretation of the Skyrme interaction
outlined above, there is no need for an underlying two-body force, 
but one can start from an effective energy functional which is 
formulated directly at the level of local one-body densities and currents.
This relaxes the fixed isotopic mix (\ref{eq:W_std}) in the
spin-orbit functional and allows more freedom for its parameterization which
was used to complement the spin-orbit interaction by an explicit 
isovector degree of freedom in the fit of the extended Skyrme functionals
SkI3 and SkI4
\begin{equation}
\label{eq:W_ext}
{\bf W}_q^{({\rm ext})}
= b_4 \; \nabla \rho + b_4' \; \nabla \rho_q
\quad .
\end{equation}
The additional isospin degree-of-freedom enables the reproduction of the 
kink in the isotope shifts of charge mean-square radii in lead, which is 
not possible with standard Skyrme forces employing (\ref{eq:W_std}) 
\cite{SkIx,Lal94a,Sha95a}, while the experimental data are reproduced by most
RMF forces. The parameters $b_4$ and $b_4'$ in SkI3 and SkI4
are adjusted to reproduce the spin-orbit splittings of protons and neutrons 
in $^{16}{\rm O}$ and the isotope shifts of charge mean-square radii in lead.
As a result of the fit the approximate relation \mbox{$b_4 \approx - b_4'$} 
emerges for SkI4, see also Table \ref{tab:SHFpar} in 
Appendix \ref{Subsect:SHF}. This means that for SkI4 the spin-orbit 
potential of one kind of nucleons depends mainly on the density
profile of the other kind of nucleons. The force SkI3 
was adjusted with the same fit strategy but with a fixed isovector 
part \mbox{$b_4'=0$} analogous to the RMF in the sense that the 
spin-orbit potentials of protons and neutrons are approximately equal.
However, there remain differences between SkI3 and the RMF:
all RMF potentials have a finite range and the spin-orbit interaction 
has a small but non--zero isospin dependence and a strong density 
dependence.
%
%
\subsection{Spin--orbit splitting}
\label{Subsect:LS:splitting}
In non-relativistic models the spin-orbit term in the 
equation-of-motion of the radial wavefunctions in case of
spherical symmetry is given by
\begin{equation}
\label{eq:Wpsi}
W_{q,r} {\textstyle \frac{1}{r}} 
\big[ j_k (j_k + 1) - \ell_k (\ell_k + 1) - {\textstyle \frac{3}{4}} 
\big] \; \phi_k (r)
\quad ,
\end{equation}
where $W_{q,r}$ is the radial component of the spin-orbit potential
and $\phi (r)$ the radial part of the single-particle wavefunction 
$\psi ({\bf r})	$.
For well-bound single-particle states, the radial wavefunctions 
$\phi_{\ell \pm 1/2}$ entering Eq.~(\ref{eq:Wpsi}) are only slightly 
different. Therefore the contributions from the potential and 
the kinetic term can be neglected in very good approximation when 
calculating the spin-orbit splitting \mbox{$\Delta \epsilon_{\rm LS} =
\epsilon_{\ell + 1/2} - \epsilon_{\ell - 1/2}$} of two states $\phi_{j}$ 
with the same radial quantum number and orbital angular momentum $\ell$ 
but different \mbox{$j = \ell \pm 1/2$}
\begin{eqnarray}
\label{eq:LSsplit}
\Delta \epsilon_{\rm LS}
& \approx &
  4 \pi
  \int \limits_0^\infty \! {\rm d}r \, r \, W_{q,r}
          \big(\ell + {\textstyle \frac{1}{2}} \big) 
  \Big[   \big(\ell + {\textstyle \frac{3}{2}} \big) 
          | \phi_{\ell + 1/2} |^2
     \nonumber \\
&  & \quad 
        - \big(\ell + {\textstyle \frac{1}{2}} \big) 
          | \phi_{\ell - 1/2} |^2
  \Big]
\quad .
\end{eqnarray} 
The spin-orbit splitting scales with $2\ell+1$ and depends sensitively 
on the overlap of the single-particle wavefunctions with $W_{q,r}/r$. 
The shape of $W_{q,r}/r$ -- which is usually peaked at the nuclear 
surface -- depends itself on the variation of the actual 
density distribution in the nucleus which changes going along
isotopic or isotonic chains, especially when the density
distribution becomes diffuse going towards the drip-lines or when
it develops a central depression -- as happens in some
superheavy nuclei, see Sect.~\ref{Subsect:120shell}.

Equation~(\ref{eq:LSsplit}) 
holds as well for the non-self-consistent single-particle
models which are used in the framework of macroscopic-microscopic
models. There the spin-orbit potential ${\bf W}$ is assumed to
be proportional to the gradient of the single-particle potential $U$. 
In the simplest case of the modified oscillator model 
-- which was used in the first studies of 
the shell structure of superheavy nuclei \cite{SuperNils,Mosel} -- 
the spin-orbit potential $W/r$ has no radial dependence, the amplitude 
of the spin-orbit splitting is simply proportional to \mbox{$2 \ell + 1$},
see \cite{Nilssonbook} for a detailed discussion. In more
refined single-particle models like the Folded-Yukawa model (FY) \cite{FY}
or Woods-Saxon model \cite{WS} the spin-orbit potential is peaked
at the nuclear surface like in the self-consistent models, 
see appendix~\ref{Subsect:FY} for details.
%
%
\section{Effective mass and average level density}
\label{Sect:effmass}
The average density of single-particle levels $g (\epsilon)$ in the vicinity 
of the Fermi energy can be estimated using the Fermi gas model in 
a finite potential well. In case of non-relativistic particles one obtains
\cite{Dob95}
\begin{equation}
\label{eq:ldSHF}
g_q^{\rm SHF} (\epsilon_{\rm F,q})
\approx \frac{3}{4} \; N_q \; \frac{2 m^\ast_q}{(\hbar k_{\rm F,q})^2}
\quad ,
\end{equation}
The relativistic generalization of formula (\ref{eq:ldSHF}) is simply
obtained by inserting the effective mass $m^{\ast}(k_F)$ at the Fermi
surface, see eq.~(\ref{eq:mstarRMF}) and the values in brackets in
table~\ref{Tab:NucMat}).

The average level density rises linearly with particle number -- the
single-particle spectra of superheavy nuclei are therefore much denser 
than those of lighter stable nuclei. This makes the
shell structure of superheavy nuclei very sensitive to details of the 
spin-orbit interaction, differences of a few $100 \; {\rm keV}$ 
in the spin-orbit splitting of two given orbitals can create or destroy 
shell closures.

The level density depends linearly on the effective mass $m^\ast$ as
well.  This causes a dramatic difference when comparing the
predictions of interactions with small effective mass, e.g.\ SkI3 with
\mbox{$m^\ast/m = 0.574$}, and parameterizations with large effective
mass like SkP with \mbox{$m^\ast/m = 1.0$} in the region of superheavy
nuclei. As said before, a phenomenological value of $m^\ast/m\approx
0.8$ for the isoscalar effective mass can be determined from the
position of the isoscalar quadrupole giant resonances which is just in
between the extremes spanned by our choice of mean field models. But a
word of caution is in place here. The value of $0.8$ is appropriate
for the effective mass in the nuclear volume. But the value may
be larger at the surface, or Fermi surface respectively
\cite{JamMah}. This is, admittedly, a feature which is not yet built
into nowadays mean field models. A thorough exploration of this aspect
is a task for future reasearch.
%
%
\section{Spherical magic numbers}
\label{Sect:magic}
%
%
\subsection{Relation of single-particle spectra and bulk properties}
\label{Subsect:SPEvsBulk}
At closed shells, one observes a sudden jump in the two-nucleon separation 
energies $S_{2q}$
\begin{equation}
S_{2q}(N_q) 
= E (N_q-2) - E (N_q) 
\quad .
\end{equation}
$N_q$ and the number of the other kind of nucleons are assumed to be even.
The two-nucleon separation energy is a better tool to quantify shell effects
than the single-nucleon separation energy due to the absence of
odd-even effects. It is a very good approximation for twice the negative 
Fermi energy
\begin{equation}
\label{eq:S2q:lambda}
S_{2q} (N,Z) 
\approx - 2 \lambda_q (N,Z)
\quad .
\end{equation}
In doubly-magic nuclei -- in which the BCS pairing model breaks down -- 
the Fermi energy is simply given by the single-particle energy of the 
last occupied state.
Deviations between the calculated and experimental values for the 
single-particle energy of the last occupied state in doubly magic 
nuclei are therefore connected by (\ref{eq:S2q:lambda}) 
with an error in the two-nucleon separation energies 
below the shell closure. Although slightly influenced by pairing correlations,
this holds in good approximation also for the first unoccupied state 
above the Fermi surface and the two-nucleon separation beyond the shell 
closure.

The size of the gap in the single-particle spectrum is given by
half the difference in Fermi energy when going from a closed
shell nucleus to a nucleus with two additional like nucleons. 
But from Eq.~(\ref{eq:S2q:lambda}) it follows that 
this is in very good approximation equal to the shell gap 
$\delta_{2q}$, the second difference of the binding energy
\begin{eqnarray}
\label{eq:d2n}
\delta_{2q} (N_q)
& = & E(N_q+2) - 2 E(N_q) + E(N_q-2)
      \nonumber \\
& \approx & - 2 [ \lambda_q (N_q+2) - \lambda_q (N_q) ] 
\quad ,
\end{eqnarray}
which was used in \cite{RutzSuper} to quantify the magicity of a nucleus.
Going away from closed shells, there is a non-negligible contribution from 
the residual pairing interaction, therefore $S_{2q}$ and $\delta_{2q}$ 
loose their direct relation to the single-particle levels. 
The two-nucleon gaps $\delta_{2q}$ represent the size of the gap in the 
single-particle spectra, but they do not contain information about 
the actual location of the single-particle energies.

Only interactions which reproduce the experimental values of the first 
single-particle state below and above the Fermi surface will give the 
correct binding energies around closed shell nuclei. This can be read the 
other way round as well: Only interactions which reproduce the binding 
energies around shell closures give a good description of at least the 
first single-particle state below and above the shell closure, but 
the bulk properties give no information on single-particle states 
away from the Fermi energy. This demonstrates nicely, however, 
that total binding energy and properties of single-particle states are 
connected in self-consistent mean-field models. This is 
very different in macroscopic-microscopic models where the bulk
properties and single-particle spectra are described in separate
models. 

One has to be careful when comparing experimental and calculated
single-particle spectra. Experimental single-particle energies
of even-even nuclei are deduced from excitation energy measurements   
of adjacent odd-mass nuclei. The binding energy of odd-mass nuclei
is affected by polarization effects induced by the odd nucleon, 
see \cite{oddNuclei} for a discussion of these effects in the framework 
of the RMF. The polarization effects are important for the comparison of 
calculated and experimental single-particle energies. But they do not
affect the relation between the single-particle spectra and the bulk
properties in even-even nuclei discussed here.
%
%
\subsection{The single-particle spectra in known nuclei}
\label{Subsect:pb208}
\label{Subsect:LSsplit:known}
\begin{figure}[t]
\centerline{\epsfig{file=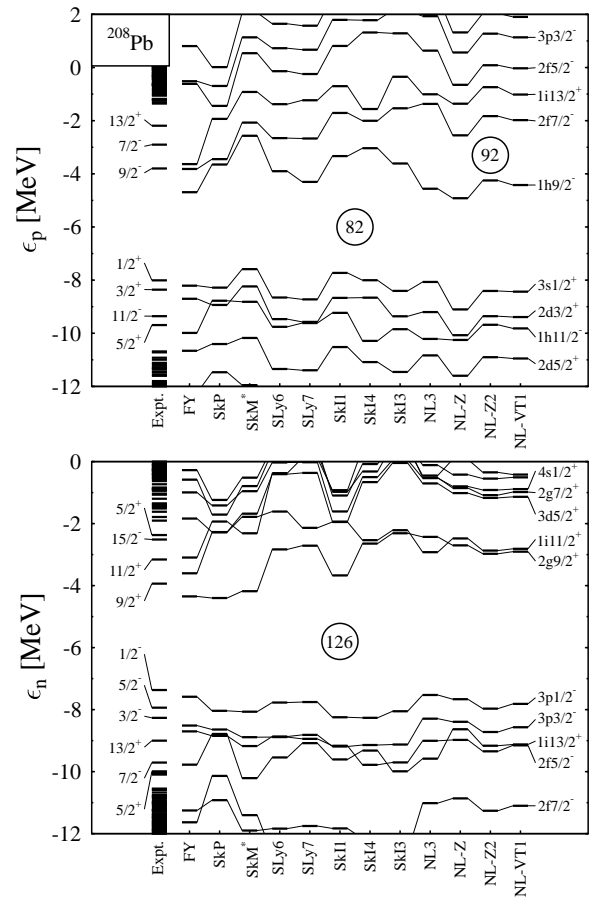}}
\caption{\label{126_082_spect}
Single--particle spectrum of the protons (upper panel) and neutrons
(lower panel) in ${}^{208}{\rm Pb}$ calculated with the mean--field 
forces as indicated.
}
\end{figure}

Before extrapolating the models to the regime of superheavy nuclei
we want to test the predictive power of the mean-field models looking 
at ${}^{208}{\rm Pb}$, the heaviest known spherical doubly-magic nucleus.
Figure~\ref{126_082_spect} shows the single-particle spectra of
${}^{208}{\rm Pb}$ as obtained from spherical calculations
with the mean-field forces as indicated. The upper panel shows 
the spectrum of the protons, the lower panel that of the neutrons.
The experimental excitation energies in the neighboring odd nuclei 
are shown on the left side for comparison, the data are taken from 
\cite{SPEexp}. The gaps in the single-particle spectra at \mbox{$Z=82$} and 
\mbox{$N=126$} are clearly visible, but the forces obviously do not
agree for this stable nucleus, which was used in the fit of all 
parameter sets employed here.
%
%
%
%

As already discussed in Sect.~\ref{Subsect:SPEvsBulk}, the difference
between the calculated and experimental energies
of the first single-particle state above and below the shell closure 
reflects the quality of the description of the total
binding energies in the vicinity of a shell closure.  
There are large differences between the forces in their predictions for
states further away from the Fermi surface. The spectrum predicted by SkP 
is much too dense and the ordering of proton states below the Fermi surface 
not reproduced. A natural explanation for this might the too large 
effective mass of SkP, but one has to be careful:
The effective mass determines the average level density only but not
the level density in an actual nucleus. The difference in energies 
between the $2g_{9/2^+}$ and $1i_{11/2^+}$ neutron states is, for example,
by far too large when calculated with SkP and SkM$^\ast$, leading to
a sub-shell closure at \mbox{$N= 136$} in contradiction to experimental
data. In the RMF and extended Skyrme forces this difference is by far 
too small, NL3 predicts even a wrong ordering of these two levels.
The relativistic forces and the relativistic corrected Skyrme force 
SkI3 overestimate the gap between the proton $1h_{9/2^-}$ and $2f_{7/2^-}$ 
states above the Fermi surface which leads to a pronounced sub-shell
closure at \mbox{$Z = 92$} which again is in contradiction with experiment.

The RMF models and the modern Skyrme forces with small effective mass 
push the $1j_{15/2^-}$ with an experimental single-particle energy
of \mbox{$\epsilon = -2.51 \; {\rm MeV}$} too much up in the spectrum, 
e.g.\ to \mbox{$\epsilon = -0.418 \; {\rm MeV}$}
in NL-VT1, while Skyrme forces with a large effective mass like 
SkM$^\ast$ and SkP work slightly better within this respect.
The differences in average level density due to the actual value
of the effective mass scale only the deviation from the experimental
value. States with large orbital angular momentum systematically lie
too high in the single-particle spectrum for all forces, see also the
proton $1i_{13/2^+}$ state. As this problem appears for all
parameterizations of both SHF and RMF models and for all nuclei
throughout the chart of nuclei \cite{Hirschegg,Gatlinburg}, we conclude that
this is not a problem of actual fits but it indicates the need for 
improved effective interactions beyond the current energy functionals.

All forces have problems to reproduce the neutron single-particle energies 
below the Fermi energy as well. All
relativistic forces and SkI3 give a wrong level ordering, the 
$2f_{5/2^-}$ state lies too low in energy in all cases. Standard Skyrme 
forces work slightly better in that respect, e.g. SkP predicts 
$2f_{5/2^-}$ to be the second-to-last state below the Fermi surface,
but interchanges the $3p_{3/2^-}$ and $1i_{13/2^+}$ states instead, 
the latter one is again pushed up too much in energy like all other 
states with large angular momentum.
It is remarkable that the non-self-consistent FY model is the
only one which reproduces the level ordering of all states
in the vicinity of the Fermi energy for both protons and neutrons.
Like the self-consistent models, however, it is not able to 
reproduce the values of the single-particle energies or even their relative 
distance.

To conclude our findings so far: the comparison between predictions
of various current mean-field models and experimental data 
shows that the models are not able to reproduce all details of 
experimental single-particle spectra and show additionally significant 
differences among each other which are related to effective mass 
and details of the spin-orbit interaction.

%
%
\begin{figure}[t]
\centerline{\epsfig{file=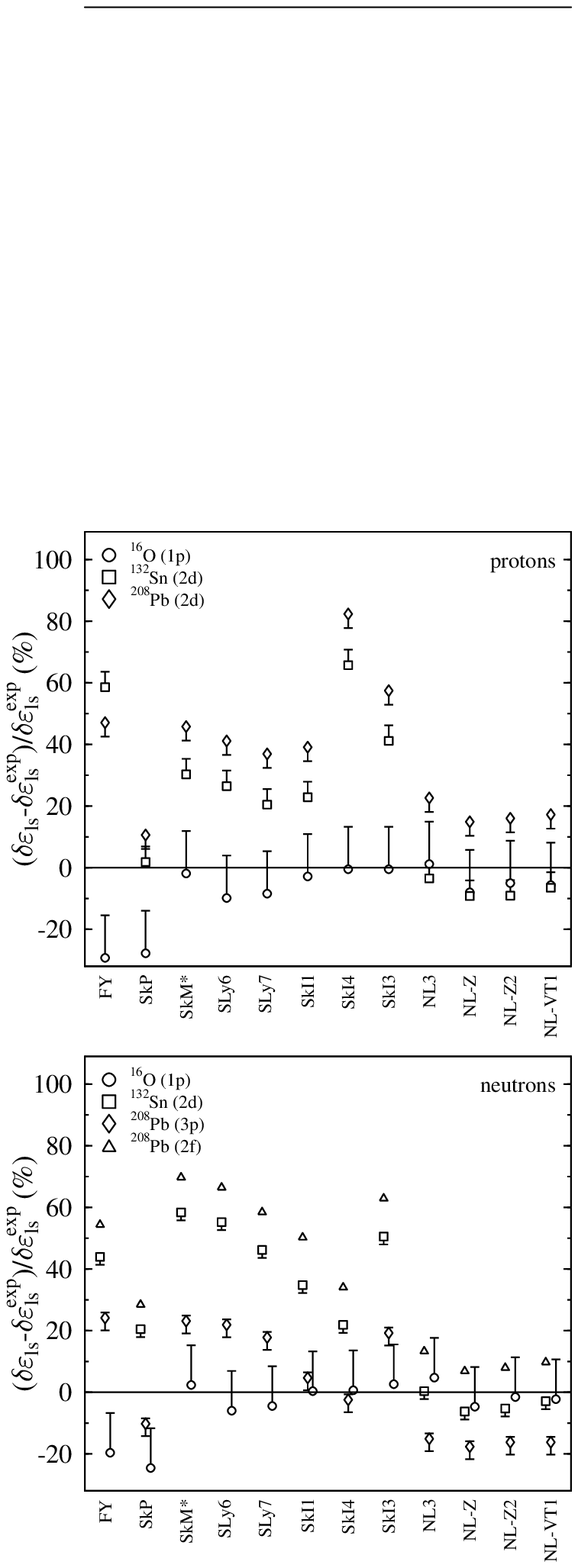}}
\caption{\label{lssplit:exp}
Relative error \mbox{$(\delta \epsilon_{ls} - \delta \epsilon_{ls}^{\rm expt})
/\delta \epsilon_{ls}^{\rm expt}$} in percent
of the spin--orbit splitting of proton 
(upper panel) and neutron (lower panel) single--particle states 
close to the Fermi surface in ${}^{16}{\rm O}$,  ${}^{132}{\rm Sn}$,
and ${}^{208}{\rm Pb}$ calculated from the mere mean--field
single--particle energies with the parameterizations as indicated.
}
\end{figure}
%
%

Shell closures of heavy nuclei are related to the spin-orbit splitting 
of states with large orbital angular momentum. Therefore it is interesting
to compare the predictions of the mean-field models
with experimental data on spin-orbit splittings in known nuclei.
Figure~\ref{lssplit:exp} shows the relative errors in $\%$ 
of the spin-orbit splittings of neutron levels (lower panel) 
and proton levels (upper panel) near 
the Fermi surface in ${}^{16}{\rm O}$, ${}^{132}{\rm Sn}$ and 
${}^{208}{\rm Pb}$. Negative errors denote theoretical values
which are too small. The spin-orbit splittings are calculated from 
the single-particle energies as they come out from a
spherical mean-field calculation. As already mentioned, the experimental 
single-particle energies are measured as separation energies between 
adjacent nuclei, where polarization effects have a visible
influence. The error bars in Fig.~\ref{lssplit:exp} represent the
uncertainty of the spin-orbit splittings due to polarization effects 
as they are found in \cite{oddNuclei}. 

All RMF forces reproduce the experimental spin-orbit splittings
fairly well, although there are deviations up to $20 \%$ which
are scattered around zero. The errors from all RMF forces are
similar and therefore it is likely that these errors represent 
the standard RMF Lagrangian, not specific parameterizations.
Although the tensor couplings of the vector mesons in NL-VT1 
change the relative distance of the single-particle energies 
compared to NL-Z2, see Fig.~\ref{126_082_spect}, they 
have no visible influence on the spin-orbit splittings 
compared to the standard Lagrangian.
It is interesting that the errors of the spin-orbit splittings 
of the neutron $3p$ and $2f$ states in ${}^{208}{\rm Pb}$ have the largest
values but different sign while ${}^{16}{\rm O}$ and ${}^{132}{\rm Sn}$
are described very well. There is only one splitting known 
for protons in ${}^{208}{\rm Pb}$ (if one excludes splittings across
the Fermi surface which have a large theoretical uncertainty, see 
\cite{oddNuclei}), so one has no information how the error depends
on the angular momentum of the state as in the case of neutrons.
But, however, the RMF gives a very good overall description of spin-orbit 
splittings throughout the chart of nuclei without any free parameters
adjusted to single-particle data.

The reproduction of the experimental data with the Skyrme functionals
is by far not as good as for the relativistic models.
There is a clear trend which is the same for all standard Skyrme forces:
for neutrons the error of the $1p$ splitting in ${}^{16}{\rm O}$ has the
smallest value, then comes the splitting of the $3p$ state in
${}^{208}{\rm Pb}$, the $2d$ state in ${}^{132}{\rm Sn}$ and then the
splitting of the $2f$ state in ${}^{208}{\rm Pb}$.
Like in the case of the RMF, the splittings of the $2f$ and $3p$ 
neutron states in ${}^{208}{\rm Pb}$ are not reproduced with the 
same quality, the error for the $2f$ state is always much larger 
compared to the experimental value than for the $3p$ state.
%
%
\begin{figure}[t!]
\centerline{\epsfig{file=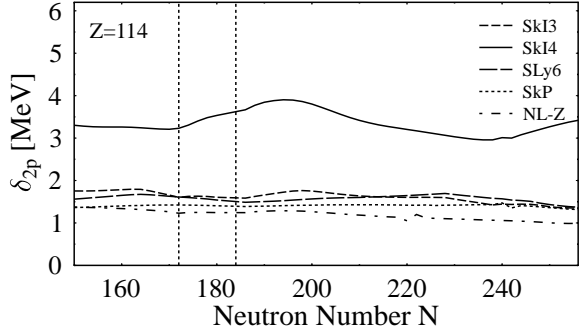}}
\caption{\label{fig:114_d2p}
Two--proton gap in the chain of \mbox{$Z = 114$} isotones calculated
with the forces as indicated.
}
\end{figure}
%
%

It is very unlucky that the parameters of the spin-orbit
interaction in non-relativistic models are usually adjusted to
data in ${}^{16}{\rm O}$, which is at the lower end of a systematic
trend increasing with mass number. Choosing one or several 
heavier nucleus for the fit, however, does not cure the
problem of the wrong trend, but it gives a better overall description
of spin-orbit splittings as can be seen from SkP, which
gives the best possible compromise for a standard Skyrme force:
the differences between the data points are similar to those from
the other standard Skyrme forces, but they are centered around zero.
The other standard Skyrme forces, SkM$^\ast$, SLy6, SLy7, and SkI1,
give similar predictions, with large errors for the $2d$ states in 
${}^{132}{\rm Sn}$ and the neutron $2f$ and proton $2d$ state in
${}^{208}{\rm Pb}$. 

The predictions of the extended Skyrme forces SkI3 and SkI4 deviate 
significantly from both the standard Skyrme forces and the RMF.  
SkI3 gives bad results 
for neutrons and protons and shows surprising large differences 
to the relativistic forces. This is somewhat unexpected because SkI3 was
constructed with the isospin dependence of the spin-orbit force which
appears in the relativistic models. This indicates that the isospin
dependence is not the only important difference between the relativistic
and non-relativistic models, density dependence or finite range of the
RMF potentials might play a much larger role for the single-particle
spectra. SkI4, gives the best results for the neutrons of all 
non-relativistic models, but at the same time it gives also the worst 
description for the proton spin-orbit splittings among all interactions
investigated here, the errors have values up to $80 \%$ for the $2d$ level 
in ${}^{208}{\rm Pb}$. The predictions for heavy nuclei might be too large
by a factor of nearly two, which makes the unique prediction of this 
force of a proton shell closure an \mbox{$Z = 114$}, caused by large 
spin-orbit splitting, not very reliable.
This will be discussed in more detail in Sect.~\ref{Subsect:114}.

The Folded-Yukawa model shows a similar behavior as the SHF forces,
but like in case of SkP the errors are scattered around zero.
%
%
\subsection{The shell structure of ${}^{298}_{184}114$}
\label{Subsect:114}
The nucleus ${}^{298}_{184}114$ is the ``traditional'' 
prediction for the spherical doubly magic superheavy nucleus 
\cite{SuperNils,Mosel,Nilssonbook} from macroscopic-microscopic models
which was confirmed in more recent models of this type 
\cite{Patyk,Mol92a,Mol94a}. As shown in \cite{Naz,RutzSuper}, most 
modern parameterizations of self-consistent models shift this property to 
larger proton numbers and/or smaller neutron numbers, depending on the 
parameterization. Only for the extended Skyrme functional SkI4 
${}^{298}_{184}114$ remains the doubly spherical magic nucleus in 
the superheavy region.

Figure~\ref{fig:114_d2p} shows the two-proton shell gap $\delta_{\rm 2p}$,
the indicator for shell closures derived from total binding energies,
for the chain of \mbox{$Z = 114$} isotopes calculated with the mean-field
forces as indicated. Only SkI4 predicts a shell closure for \mbox{$Z = 114$}, 
all other forces give rather small $\delta_{\rm 2p}$.
In contrast to the proton shell closures at higher charge numbers $Z$
which will be discussed in the following,
the \mbox{$Z = 114$} shell is stable for varying neutron number.
%
%
\begin{figure}[t!]
\centerline{\epsfig{file=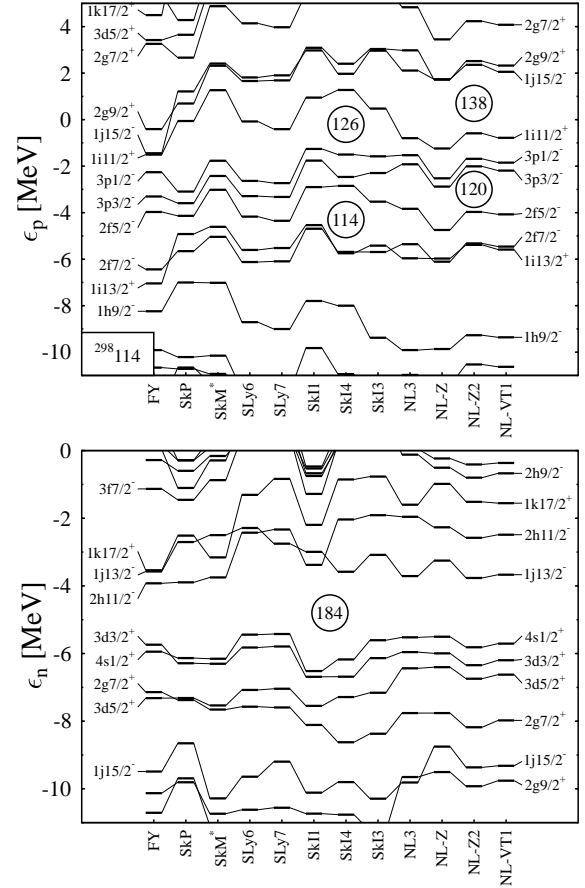}}
\caption{\label{184_114_spect}
Single--particle spectra of ${}^{298}_{184}114$ for protons (top) 
and neutrons (bottom) at spherical shape for the mean--field forces 
as indicated.
}
\end{figure}
%
%

We want to see now how the different predictions for the shell gap
$\delta_{\rm 2p}$ in the potential doubly-magic nucleus ${}^{298}114$ 
are reflected in its single-particle spectra, see Figure~\ref{184_114_spect}.
The possible shell closure at  \mbox{$Z = 114$} is located between two 
spin-orbit coupled states, the $2f_{7/2^-}$ and $2f_{5/2^-}$ levels.
Additionally, the $1i_{13/2^+}$ state which has a similar energy
as the $2f$ states has to be pushed down. Therefore it is 
immediately clear that \mbox{$Z = 114$} is only magic in 
the case of a large amplitude of the spin-orbit splitting.
A strong \mbox{$Z = 114$} shell appears only for SkI4, the force 
with the largest proton spin-orbit splitting in this nucleus of all forces 
under investigation. But it is to be remembered that SkI4 overestimates 
the spin-orbit splitting of the protons in ${}^{208}{\rm Pb}$ by $80 \%$.
This makes the prediction of a large spin-orbit splitting in
${}^{298}114$, leading to a strong shell closure, very doubtful.

SkP, the force with effective mass \mbox{$m^\ast/m = 1.0$}
and therefore a large density of single-particle levels shows no 
significant shell structure at the Fermi surface of the protons at all. 
For all other forces there is at least a sub-shell closure at \mbox{$Z = 114$}.
But only for SkI4 the gap in the single-proton spectrum
is large enough to be interpreted as a major shell closure.
For all standard Skyrme forces the  $1i_{13/2^+}$ state is located between
the $2f$ states, which significantly reduces the  \mbox{$Z = 114$} gap.

In some of the other forces with smaller spin-orbit splitting, like
SkI3 and the RMF parameterizations, there is a gap in the spectrum
at \mbox{$Z = 120$} indicating the major shell closure 
of these forces, while in all Skyrme forces there appears 
a gap at \mbox{$Z = 126$}, hinting at another potential spherical magic 
proton number. But as we will see in what follows the gap at \mbox{$Z = 126$}
becomes smaller with increasing proton number and has disappeared for 
most of the forces when reaching this proton number.
%
%
\begin{figure}[b!]
\centerline{\epsfig{file=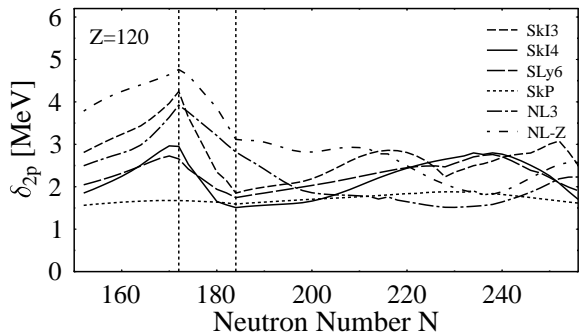}}
\caption{\label{fig:120_d2p}
Two--proton gap in the chain of \mbox{$Z = 120$} isotones calculated
with the parameterizations as indicated. 
}
\end{figure}
%
%
In the single-particle spectrum for the neutrons in ${}^{298}_{184}114$ 
the differences between the various mean-field forces are much smaller 
than for the protons. All forces show a gap in the single-neutron spectrum at 
\mbox{$N = 184$}, but for the relativistic parameterizations 
the amplitude of this gap is smaller than for the Skyrme forces and even
decreases with increasing effective mass. Therefore, in NL3
(the RMF force with the largest effective mass)
the major shell closure at \mbox{$N = 184$} has vanished. 

The single-particle spectra of both protons and neutrons
from the non-self-consistent FY model
look very different compared to all self-consistent models.
In particular, the spin-orbit splitting of all proton states is much larger
compared to all self-consistent models with the exception of SkI4.
At the Fermi surface, the $1i_{13/2^+}$ proton state which is the last filled 
state in all standard Skyrme forces, is pushed down below the
$2f_{7/2^-}$ state by the large spin-orbit splitting. 
This creates the large gap in the single-particle spectrum at 
\mbox{$Z = 114$}. 

Although the non-self-consistent FY model predicts \mbox{$N = 184$}
to be magic as well, the ordering of the neutron states below
the \mbox{$N = 184$} shell closure is very different.
The large spin-orbit splitting in the FY model pushes the $3d_{3/2^+}$
state above the $4s_{1/2^+}$ state and the $3d_{5/2^+}$ below the
$2g_{7/2^+}$ state.
Another difference to the self-consistent models is the large level
density above the gap at \mbox{$N = 184$}. Three states with large
angular momentum, i.e.\ $2h_{11/2^-}$, $1j_{13/2^-}$ and $1k_{17/2^+}$
are close together which explains that the maximum of the corresponding
shell correction is shifted to nuclei with the somewhat smaller 
(and non-magic!) neutron number around \mbox{$N = 178$} \cite{Mol94a}.
%
%
\subsection{The \mbox{$Z = 120$} shell}
\label{Subsect:120shell}
%
%
\begin{figure}[t!]
\centerline{\epsfig{file=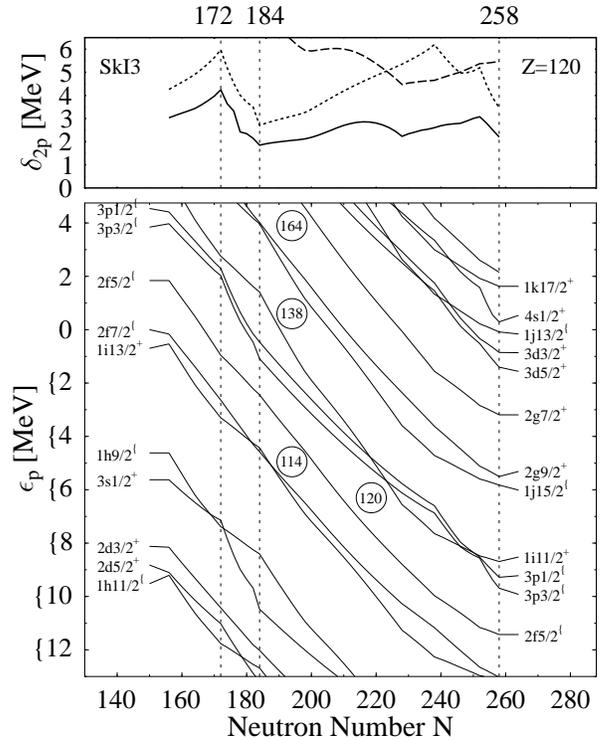}}
\caption{\label{z120_pspec_ski3}
Single--proton levels in the vicinity of the Fermi
energy for the isotopes of \mbox{$Z = 120$} (lower panel)
and two--proton shell gap $\delta_{\rm 2p}$ (upper panel)
versus the neutron number, computed with SkI3.
The dotted line in the upper panel is twice the difference between
the $3p_{1/2^-}$ and $2f_{5/2^-}$ levels, the dashed line twice the 
difference between the $1i_{11/2^+}$ and $2f_{5/2^-}$ levels.
}
\end{figure}
%
%
In self-consistent models, the occurrence of a 
spherical proton shell closure with
given $Z$ can change with varying neutron number $N$, and similarly
the neutron shell closures can vary with changing proton numbers. 
While for light nuclei this happens only at the limits of stability, 
e.g.\ the vanishing of the \mbox{$N = 28$} shell for proton numbers 
\mbox{$Z < 20$} which is hinted experimentally \cite{Lew89,Sor93,Sch96,Gla97} 
and predicted by self-consistent mean-field models \cite{Wer96a,Lal98a}.
In the region of superheavy nuclei the nucleon number dependence of shell
closures is a common feature in the predictions of self-consistent models
\cite{RutzSuper,BuervenSuper}.
%
%
\begin{figure}[t!]
\centerline{\epsfig{file=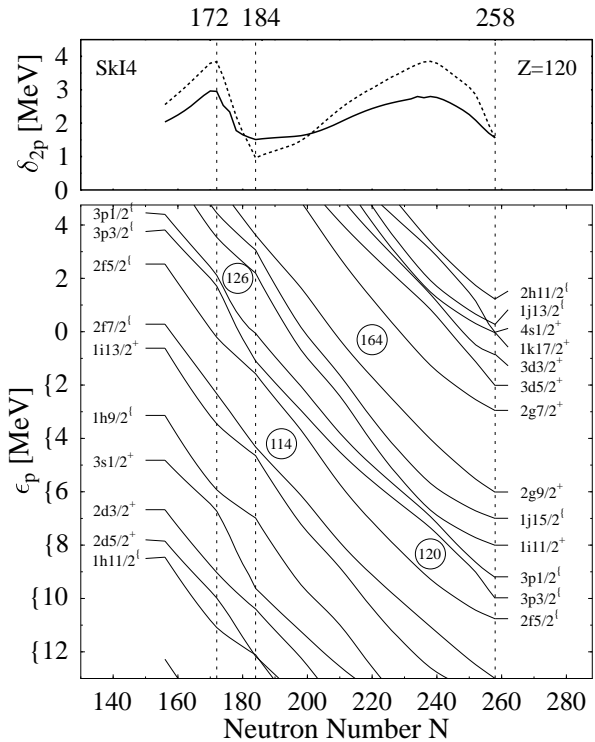}}
\caption{\label{z120_pspec_ski4}
The same as in Fig.~\protect\ref{z120_pspec_ski3},
but computed with SkI4.}
\end{figure}
%
%

The most important example is the spherical \mbox{$Z = 120$} shell, see 
Fig.~\ref{fig:120_d2p} which shows the two-proton shell gap 
$\delta_{\rm 2p}$ of the \mbox{$Z = 120$} isotones for some of the 
forces under investigation. 
All parametrizations except SkM$^\ast$ and SkP predict a peak in the 
$\delta_{\rm 2p}$ at \mbox{$N = 172$} which is followed by a steep decrease 
of $\delta_{\rm 2p}$ when going towards larger neutron numbers. 
The  $\delta_{\rm 2p}$ are largest in the relativistic parametrizations and
the extended Skyrme functional SkI3 with the RMF-like spin-orbit interaction,
but even most of the standard Skyrme forces, i.e.\ those with small
effective mass, show an enhanced $\delta_{\rm 2p}$ around \mbox{$N = 172$} 
as well.

To understand the origin of the neutron-number dependence,
Fig.~\ref{z120_pspec_ski3} shows the single-proton spectra (lower panel)
and the corresponding $\delta_{\rm 2p}$ (upper panel) of the
\mbox{$Z = 120$} isotones calculated with SkI3.
The quantity of interest is the gap in the spectrum at \mbox{$Z = 120$}.
First of all it is to be noted that the single-particle spectrum is indeed 
relatively dense. Therefore already minimal relative changes of the 
proton levels produce a regime of higher level density at the proton Fermi
surface around \mbox{$N = 184$}, the neutron number where the proton shell
gap is lowest. The relative changes of the levels are due to changes in the 
amplitude of the spin-orbit splitting. The shell closure at \mbox{$Z = 120$} 
can appear only when the spin-orbit splitting between the $2f$ 
proton states below the Fermi energy and the $3p$ 
states above the Fermi energy is small. In nuclei for which the 
spin-orbit splitting of these levels is large, e.g.\ around 
\mbox{$N = 184$}, the gap in the single-particle spectrum at \mbox{$Z = 120$}
vanishes. 
%
%
\begin{figure}[t!]
\centerline{\epsfig{file=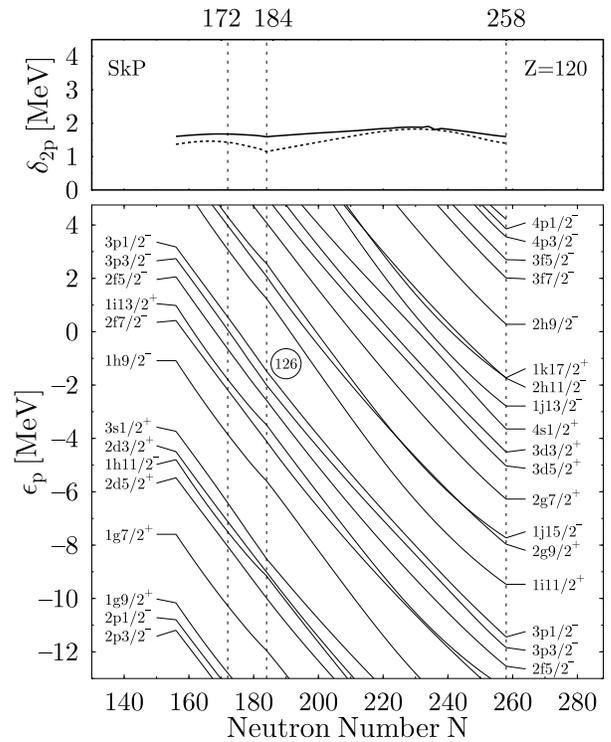}}
\caption{\label{z120_pspec_skp}
The same as in Fig.~\protect\ref{z120_pspec_ski3},
but computed with SkP.}
\end{figure}
%
%

To demonstrate the relation between the shell gap calculated from total 
binding energies and the actual gap in the single-particle spectrum, in
the upper panel of Fig.~\ref{z120_pspec_ski3} the difference in energy 
$\Delta \epsilon$ between the last single-particle
state below and the first state above the Fermi energy is shown with
a dotted line. As can be clearly seen, $\Delta \epsilon$ is always
larger than $\delta_{\rm 2p}$, showing that the shell gaps $\delta_{\rm 2p}$
calculated from total binding energies are influenced by the pairing,
which smears out the shell effects.

For SkI4 the spin-orbit splitting of the single-proton levels
in superheavy nuclei is in general larger than for SkI3,
see Fig.~\ref{z120_pspec_ski4}. Therefore the magic number \mbox{$Z = 114$}
appears, corresponding to a large gap between the $2f$ 
single-proton levels. Like for SkI3, the spin-orbit splitting of the
levels in the vicinity of the Fermi energy is largest around 
\mbox{$N = 184$}. While this effect weakens the shell gap at 
\mbox{$Z = 120$} in SkI3 and SkI4, it amplifies the gap in the single-proton
spectrum at \mbox{$Z = 114$} in SkI4. The magic \mbox{$Z = 120$} appears for
SkI4 only for isotopes with relatively small spin-orbit splitting
in the vicinity of the Fermi energy, i.e.\ at large neutron numbers.
%
%
\begin{figure}[t!]
\centerline{\epsfig{file=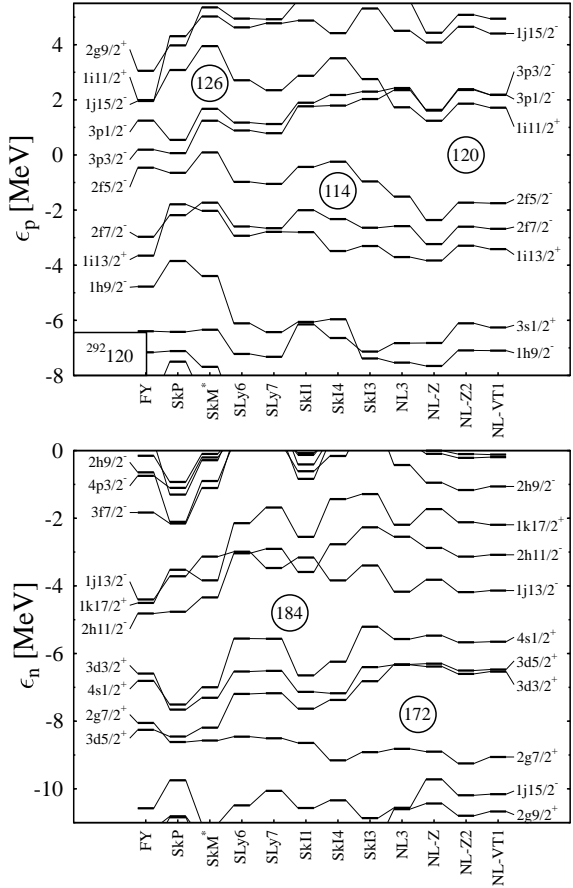}}
\caption{\label{172_120_spect}
The same as in Fig.~\protect\ref{184_114_spect},
but for ${}^{292}_{172}120$}
\end{figure}
%
%

The single-particle spectra of the protons look very different for
forces with large effective mass, e.g.\ SkP, see Fig.~\ref{z120_pspec_skp}.
Owing to the large average level density at the Fermi surface there are no
distinct shell effects at all for the \mbox{$Z = 120$} isotopes. 
Additionally, there are only slight changes of the level structure
with varying neutron number $N$. This
confirms our previous finding that a large effective mass washes
out most of the shell structure in superheavy nuclei. In this case, 
the proton shell gap $\delta_{\rm 2p}$ and the last single-particle 
level below the Fermi energy and the first level above are in good 
agreement.
%
%
\subsection{The Shell Structure of ${}^{292}_{172}120$}
\label{Subsect:172120}
The occurrence of the proton shell closure at \mbox{$Z = 120$} is
coupled to at least a subshell closure at \mbox{$N = 172$}. Therefore
it is interesting to take a detailed look into the single-particle spectra 
of ${}^{292}_{172}120$, which are shown in Fig.~\ref{172_120_spect}.
The upper panel shows the proton levels, the lower one shows 
the neutron levels. As already discussed in Sect.~\ref{Subsect:120shell},
the occurence of the shell closure at \mbox{$Z = 120$} depends on
the amplitude of the spin-orbit splitting of the $3p$ states 
above the Fermi level and the $2f$ levels below the Fermi energy. It
appears only when the level density at the Fermi energy is small and
the spin-orbit splitting is weak, but this is the 
case for all forces under investigation except SkP and SkM*, the 
forces with the largest effective mass and therefore largest (average)
level density. It has to be noted that for almost all forces this 
nucleus is located near the two-proton drip line since the first
unoccupied proton level has a positive single-particle energy.
%
%
\begin{figure}[t1]
\centerline{\epsfig{file=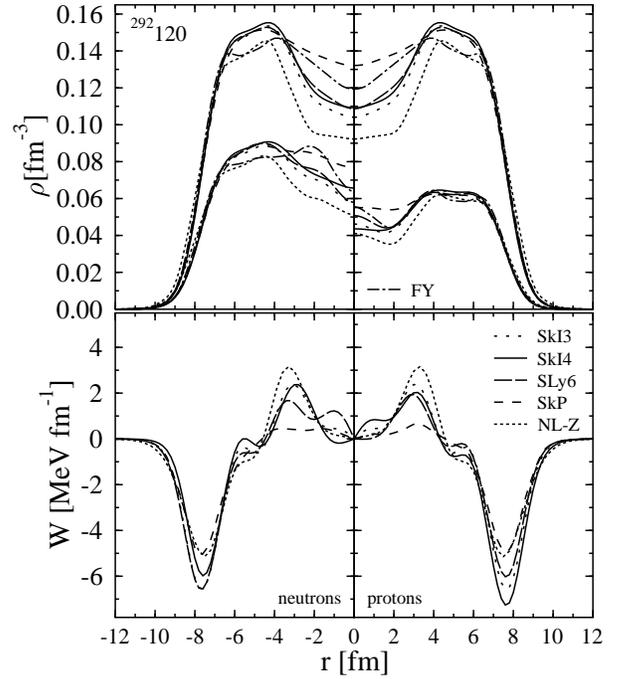}}
\caption{\label{172120rho}
Density distribution (upper panel) and radial component of the 
spin--orbit potential (lower panel) of protons (right), neutrons 
(left) for ${}^{292}_{172}120$, calculated with the forces a indicated.
The total density is plotted in the upper panels as well.
The density distributions calculated from the single-particle wavefunctions 
as they come out in the FY model is drawn for comparison. All models
except SkP show a central depression in the density distribution, which
has a visible impact on the spin-orbit potential.
}
\end{figure}
%
%
%
%
\widetext
\begin{figure*}[t!]
\centerline{\epsfig{file=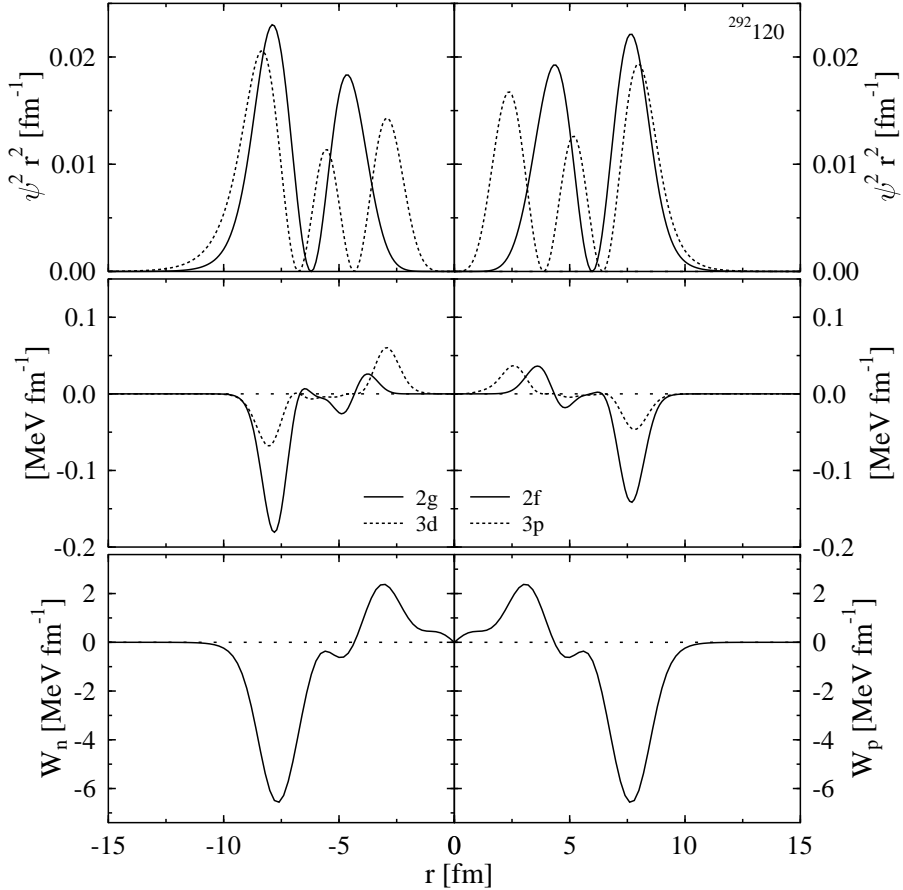}}
\caption{\label{172120W:SkI3}
Radial density distribution (upper panel), integral kernel of
the spin--orbit splitting (\protect\ref{eq:LSsplit}) (middle panel)
and radial component of the spin--orbit potential $W_r$ for the
$2f$ and $3p$ proton states (right) and $2g$ and $3d$ neutron states
in ${}^{292}_{172}120$, calculated with SkI3. The probability distribution 
is shown for the state with larger total angular momentum only.
}
\end{figure*}
\narrowtext
%
%

The level ordering of the proton states above the Fermi level for the RMF 
forces NL-Z, NL-Z2 and NL-VT1 is quite unusual, the $3p$ state with
small total angular momentum is located \emph{above} the state with large 
angular momentum. This phenomenon is related to the unusual shape of the 
density distribution of this nucleus, see the upper panel of 
Fig.~\ref{172120rho}. The large dip at the nuclear center, where the density 
is reduced to $2/3$ of its nuclear matter value, leads to a region 
around \mbox{$r \approx 3 \; {\rm fm}$} where the spin-orbit potential 
has the opposite sign, see the lower panel of Fig.~\ref{172120rho}. 
Therefore, for $j$
states with large occupation probability in this region the amplitude
of the spin-orbit splitting is dramatically reduced or even has the
opposite sign as it is the case for NL-Z, NL-Z2 and NL-VT1. 
Additionally, this density
distribution strongly affects the shape of the single-particle
potentials, which are reduced at the nuclear center by approximately
the same factor as the density.  Orbitals with large angular momentum,
e.g.\ the $1i$ states, are pushed down in the spectrum compared to
states with rather small angular momentum like the $3p$ states. This
leads to a completely different level ordering above the \mbox{$Z =
120$} proton shell in case of the RMF forces.

The same effect occurs in the neutron spectrum as well.
The level ordering of the $3d$ states is reversed for the RMF forces, 
see the lower panel of Fig.~\ref{172_120_spect}. Again, for SkP, the force 
which gives the less pronounced dip of the density distribution, the 
spin-orbit splitting of the $3d$ neutron states is largest.
States with large angular momentum and therefore small overlap with the 
center of the nucleus, i.e.\ the $2g$ or $1j$ states, show the common 
spin-orbit splitting.
%
%
\mediumtext
\begin{figure*}[t!]
\centerline{\epsfig{file=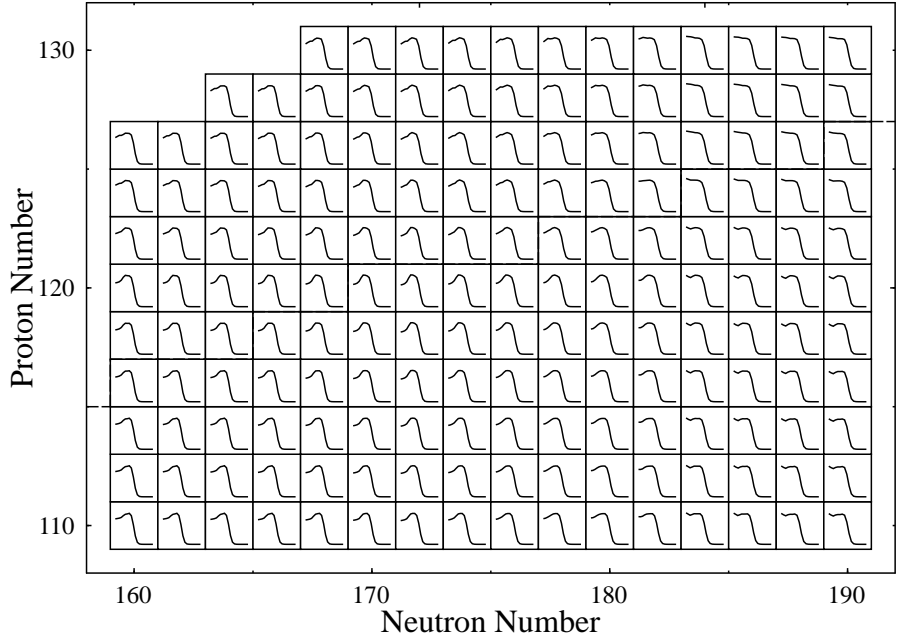}}
\caption{\label{ski3_shapes}
Distribution of the mass density from spherical calculations with
SkI3 in the region of the \mbox{$Z=120$}, \mbox{$N=172$}, and 
\mbox{$N=184$} shells.
}
\end{figure*}
\narrowtext
%
%

The details of this effect as they appear in the non-relativistic
SkI3 are shown quantitatively in Fig.~\ref{172120W:SkI3} for selected
neutron (left) and proton states (right), in both cases one level with large 
and one with small orbital angular momentum close to the Fermi 
energy. The upper panels show the radial density distributions
$4 \pi \, r^2 \, |\phi(r)|$ of the $2g$ and $3d$ neutron states and 
$2f$ and $3p$ proton states, where $\phi(r)$ is the radial component of the 
single-particle wavefunction $\psi (\bf{r})$. The
radial density is shown for the state with larger
total angular momentum only. The middle panels shows the integrand 
$ r \, W_{q,r} \, (\ell + {\textstyle \frac{1}{2}} ) 
                  [ (\ell + {\textstyle \frac{3}{2}} ) | \phi_{\ell + 1/2} |^2
                  - (\ell - {\textstyle \frac{1}{2}} ) | \phi_{\ell - 1/2} |^2]
$
which enters the calculation of the spin-orbit splitting
(\ref{eq:LSsplit}), while the radial component of the spin-orbit
potential $W_r$ is shown in the lower panels. Besides the familiar
attractive peak at the surface of the nucleus, the central depression of 
the density leads to a repulsive peak of the spin-orbit potential around
\mbox{$r \approx 3 \; {\rm fm}$}. The total spin-orbit splitting now depends
sensitively on the location of the radial wavefunctions. The neutron
$3d$ and proton $3p$ states with three nodes but small angular momentum
have large overlap with both the repulsive
and the attractive part of the spin-orbit potential (note that small
radii are supressed only with $1/r$ and not as usual with $1/r^2$),
leading to nearly vanishing spin-orbit splitting, while the neutron
$2g$ and proton $2f$ states with only two nodes feel only the 
spin-orbit potential at the nuclear surface (and have much larger
overlap with this than the small-angular-momentum states), 
showing the usual spin-orbit splitting.

Note that this is a polarization effect that is naturally 
included in the self-consistent description of nuclei but cannot occur in 
semi-microscopic approaches like the ``Extended Thomas-Fermi-Strutinski 
Integral'' method (ETFSI) \cite{Pea91a,ETFSI} or macroscopic-microscopic 
models \cite{FRDM} with prescribed densities and/or single-particle potentials,
where one has a very restricted variational freedom of the density 
profile only (ETFSI) or no degree-of-freedom in the density distribution
and single-particle potentials at all (macroscopic-microscopic).
Looking at the spectrum calculated with the FY model, the spin-orbit
splitting is indeed much larger than in self-consistent models, 
especially for the $3p$ proton and $2g$ neutron states which are crucial for
the \mbox{$N = 172$} shell closure. Comparing Fig.~\ref{172_120_spect} with 
Fig.~\ref{184_114_spect} for ${}^{298}{114}$ one immediatley sees that 
the change in the single-particle spectra of both protons and neutrons
predicted by FY is much smaller when going from ${}^{298}{114}$ 
to ${}^{292}{120}$ than in all self-consistent models.

Figure~\ref{ski3_shapes} shows the profile of the total density 
in even-even nuclei in the region of 
the \mbox{$Z = 120$}, \mbox{$N = 172$} and \mbox{$N = 184$} shells
as they result from spherical calculations with SkI3. 
This demonstrates that the density profiles are coupled to the 
shell closures (and vice versa). At large neutron numbers \mbox{$N > 184$}
all nuclei have the usual density profiles, while going below 
\mbox{$N = 184$} the nuclei immeadiately show a central depression
that is most pronounced for nuclei with \mbox{$Z = 120$}.
It is noteworthy from Fig.~\ref{ski3_shapes} 
that the central depression of the density distribution is
coupled to the neutron number -- it disappears for all neutron numbers
above \mbox{$N = 184$}, while the density profiles of nuclei with 
constant neutron number but different proton number look very similar.
The reason for this is that the last filled neutron levels below the
\mbox{$N = 172$} gap -- $2g_{9/2^+}$ and $1j_{15/2^-}$ and $2g_{7/2^+}$ --
all have large orbital angular momentum and are therefore mainly located 
at the nuclear surface. Going from \mbox{$N = 172$} to \mbox{$N = 184$}
only levels with small angular momentum -- $3d_{5/2^+}$, $3d_{3/2^+}$ and 
$4s_{1/2^+}$ -- are occupied which have a large probability distribution 
at small radii. This means that the unusual density distribution 
of nuclei around ${}^{292}_{172}120$ is simply caused by the filling 
of the neutron levels which have the same ordering in all models 
investigated here. This effect thus should occur in non-self-consistent
models as well. And indeed the densities calculated from the FY model 
(plotted in the upper panel of Fig.~\ref{172120rho}) 
show the same behaviour as the densities from 
the self-consistent models, although the effect is weaker here.
But unlike the non-self-consistent models with prescribed
potentials, the densities in self-consistent models are fed back into
the potentials which amplifies the effect by driving the wavefunctions 
to larger radii. Additionally the self-consistent spin-orbit potentials 
are influenced which in turn causes the \mbox{$Z = 120$} proton 
shell closure.
%
%
\begin{figure}[t!]
\centerline{\epsfig{file=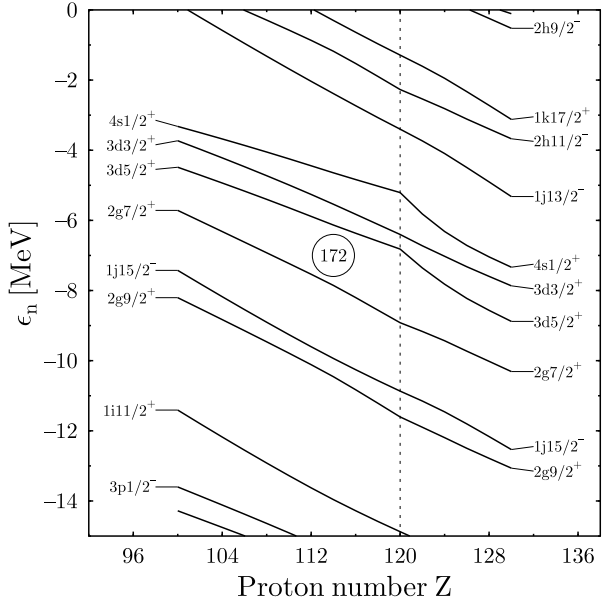}}
\caption{\label{n172_nspec_ski3}1
Single--neutron levels in the vicinity of the Fermi
energy for the isotones of \mbox{$N = 172$} versus the
proton number, computed with SkI3.
}
\end{figure}
%
%

The same effect which creates the \mbox{$Z = 120$} proton shell is
responsible for the appearance of a magic neutron number \mbox{$N = 172$}.
The gap at \mbox{$N = 172$} depends sensitively on the amplitude 
of the spin-orbit splitting of the $3d$ neutron levels above this gap.
Therefore it occurs again only for the RMF parametrisations and the
generalized Skyrme functional SkI3. It can be expected that this neutron 
shell closure is restricted to nuclei with a prominent central depression 
of the density
like the \mbox{$Z = 120$} proton shell closure. Fig.~\ref{n172_nspec_ski3}
shows the single-particle energies of the neutrons in the chain
of \mbox{$N = 172$} isotones calculated with SkI3. The \mbox{$N = 172$} gap
is largest for \mbox{$Z = 120$} in agreement with our findings for
the $\delta_{\rm 2n}$ in \cite{RutzSuper}. Although all these \mbox{$N = 172$} 
isotones show a central depression of the density distribution, 
for those those around \mbox{$Z = 120$} the decrease in density 
when going to small radii is steepest. This gives the largest
(positive) peak in the spin-orbit potential and therefore the
smallest spin-orbit splitting of the neutron $3d$ levels which in turn
gives the largest gap in the spectrum.
%
%
\subsection{The Shell Structure of ${}^{310}_{184}126$}
%
%
\begin{figure}[b!]
\centerline{\epsfig{file=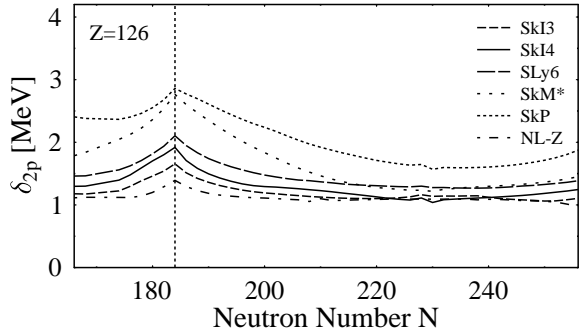}}
\caption{\label{fig:126_d2p}
Two--proton gap in the chain of \mbox{$Z = 126$} isotones calculated
with the parameterizations as indicated.
}
\end{figure}
%
%
The question weather \mbox{$Z = 114$} or \mbox{$Z = 126$} is the next
spherical shell closure beyond the experimentally known \mbox{$Z = 82$} 
is as old as the first extrapolations of nuclear shell structure to 
superheavy nuclei in simple models. While 
\mbox{$Z = 126$} corresponds to the largest experimentally known magic 
neutron number, \mbox{$Z = 114$} has no counterpart for the neutrons. A 
large number of self-consistent models predict \mbox{$Z = 120$} 
to be the next proton shell closure, but there are some parametrisations 
predicting \mbox{$Z = 126$} as an alternative.

Figure~\ref{fig:126_d2p} shows the two-proton shell gap $\delta_{\rm 2p}$
for the chain of \mbox{$Z = 126$} isotopes calculated with the forces as 
indicated. For SkP and SkM$^\ast$ two Skyrme forces forces which both have 
a large effective mass this is a major spherical shell closure. As in case 
of \mbox{$Z = 120$} the shell closure is neutron-number dependent,
it fades away when going to neutron numbers beyond \mbox{$N = 184$}.
For most other Skyrme forces there is only a slight enhancement of 
$\delta_{\rm 2p}$ in a small vicinity around \mbox{$N = 184$} which
cannot be interpreted as a shell closure. The forces with ``relativistic'' 
spin-orbit coupling, i.e.\ all RMF forces and SkI3, predict very small 
shell gaps only.
%
%
\begin{figure}[t!]
\centerline{\epsfig{file=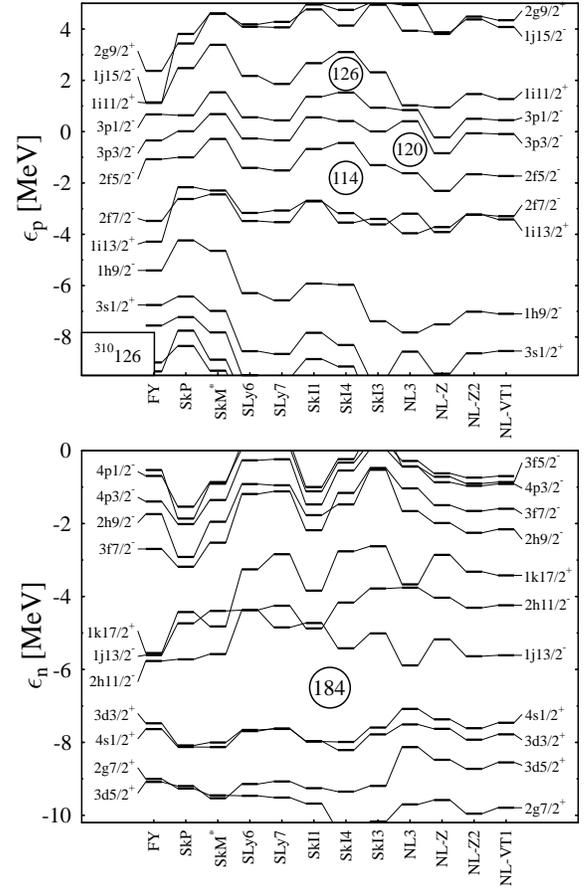}}
\caption{\label{184_126_spect}
The same as in Fig.~\protect\ref{184_114_spect},
but for ${}^{310}_{184}126$.
}
\end{figure}
%
%

This is reflected in the single-particle spectra, see Fig.~\ref{184_126_spect}.
Contrary to the appearance of the \mbox{$Z = 114$} and \mbox{$Z = 120$} 
shell closures, which can be explained simply by looking
at the spin-orbit splitting of adjacent proton levels, the situation 
is more complicated for the \mbox{$Z = 126$} shell closure.
The $1i_{11/2^+}$ proton state which lies above the \mbox{$Z = 126$} gap 
is widely separated from the deeply bound $1i_{13/2^+}$ state. 
Therefore the appearance of the magic number \mbox{$Z = 126$} 
does not depend only on the amplitude of spin-orbit splitting but on the
relative distance of levels with different orbital angular momentum as well,
although all relativistic forces with overall small spin-orbit splitting 
show no shell closure at \mbox{$Z = 126$}. Remembering that states with 
large angular momentum have systematically too small single-particle 
energies and that the spin-orbit splitting predicted by the standard 
Skyrme forces and SkI4 is too large in heavy nuclei -- both would
reduce the \mbox{$Z = 126$} gap -- the occurrence of a proton shell 
closure at \mbox{$Z = 126$} is very questionable.

Comparing the single-proton spectra of ${}^{298}_{184}114$ 
(Fig.~\ref{184_114_spect}) and ${}^{310}_{184}126$ one sees immedeatly 
that the gap at \mbox{$Z = 126$} 
becomes much smaller with increasing proton number. An exception is 
the non-selfconsistent FY model, here the relative distances of all proton 
and neutron have only slightly changed. 
This gives a further example for the strong dependence of the 
shell structure of superheavy nuclei on the nucleon numbers in
self-consistent models. 

For all forces the Fermi energy is positive which means that 
${}^{310}_{184}126$ is predicted to be unstable against proton emission. 
However, owing to the large Coulomb barrier in superheavy nuclei we 
expect that this nucleus decays trough other more common channels.
%
%
\subsection{Spin-orbit splitting in superheavy nuclei}
\label{Subsect:LSsplitting}

\begin{figure}[t!]
\centerline{\epsfig{file=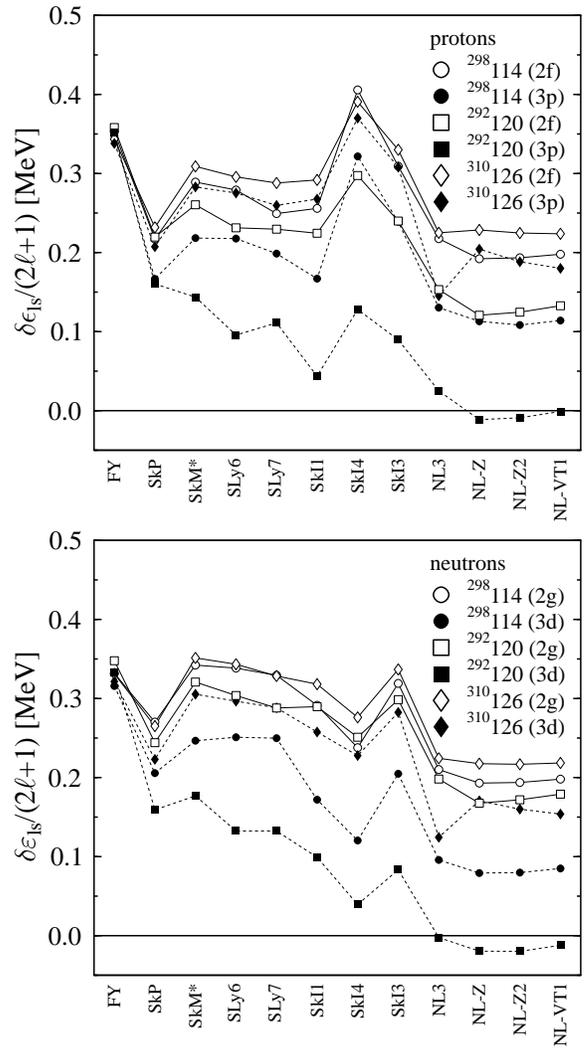}}
\caption{\label{lssplit:sh}
Amplitude of the spin--orbit splitting of several superheavy
nuclei as predicted by the mean--field forces as indicated.
The ls-splitting is weighted with $1/(2\ell+1)$ to remove the
trivial dependency on the orbital angular momentum. This shows nicely
that in self--consistent models the ls-splitting has an additional
state--dependency that does not occur in simple potential models
like FY (In the modified oscillator model the splitting is simply
the $\kappa$ parameter in the potential) that is related to the
shape of the density distribution.
}
\end{figure}

We have seen that the predictions of self-consistent models for the
spin-orbit splitting in superheavy nuclei shows a pronounced dependence
on the nucleon numbers and the orbital
angular momentum of the single-particle states. This is summarized
in Fig~\ref{lssplit:sh}. The upper panel shows the spin-orbit
splitting of the $3f$ (white markers) and $3p$ (black markers) 
proton states, while the lower panel shows the splitting of the 
$2g$ (white markers) and $3d$ (black markers) neutron states
in the nuclei as indicated for all forces under investigation.
The trivial trend with the orbital angular momentum $\ell$ of the 
states is removed dividing by $2 \ell + 1$, see Eq.~(\ref{eq:LSsplit}).

While in the non-self-consistent Folded-Yukawa (FY) model all 
states have nearly the same renormalized spin-orbit splitting, 
there are large differences between the self-consistent models. 
The predictions of the forces for certain states in certain nuclei
differ as such, but there are clearly visible trends with nucleon
number and orbital angular momentum which occur for all 
parameterizations. Picking out one force, one sees in most cases 
the same pattern: 
The spin-orbit splitting of a given state in ${}^{310}_{184}126$ 
is larger than in ${}^{298}_{184}114$, while it is smallest in
${}^{292}_{172}120$. The (renormalized) splitting of states with large 
orbital angular momentum is always larger than the splitting of states 
with small orbital angular momentum. As already discussed above, this 
is related to the shape of the nuclear density distribution and the
effect is largest in ${}^{292}_{172}120$, for which most self-consistent
forces predict a pronounced central depression in the density.
%
%
%
%

There is a difference between protons and neutrons. While the splitting
of the $2g$ neutron state is comparable in all nuclei (although it 
follows the trend mentioned above), the differences with mass number 
for the $2f$ proton states is much more pronounced.

There are large differences between the various forces. The 
parameterizations can be divided into three groups which differ in
the isospin dependence of the spin-orbit interaction: standard Skyrme
forces (SkP-SkI1), extended Skyrme forces (SkI3, SkI4) and RMF forces
(NL3, NL-Z, NL-Z2, and NL-VT1). The standard Skyrme forces
in most cases  predict larger spin-orbit splittings than the RMF forces. 
Like in the case of the known nuclei, the predictions of the extended
Skyrme forces SkI3 and SkI4 do not stay in between the predictions of
standard Skyrme forces and the RMF. Again, SkI3 gives much larger 
spin-orbit splittings than the RMF forces with a similar isospin 
dependence of the spin-orbit interaction, while SkI4 stays in betwen 
standard Skyrme forces and RMF forces for neutrons, but gives the largest 
splittings for proton levels.
For SkP, the force with large effective mass \mbox{$m^\ast/m = 1.0$}
and the smallest spin-orbit parameter $b_4$, the results look 
somewhat different as it was already the case for the known nuclei
discussed in Sect.~\ref{Subsect:LSsplit:known}, the spin-orbit splitting
of the large angular momentum states and the dependence of the amplitude 
of the renormalized spin-orbit splitting on the orbital angular momentum
is smaller than in other standard Skyrme forces.

The predictions for shell closures are sensitive on the isospin 
dependence of the spin-orbit interaction and the isoscalar 
effective mass. But there are additional dependencies of the 
spin-orbit splitting than the already mentioned ones as can be seen by
comparing SkI3 and the RMF forces, which have similar effective mass and
isospin dependence of the spin-orbit interaction.
%
%
\section{Summary and conclusions}
\label{Sect:summary}
We have investigated the influence of the isospin dependence of the
spin-orbit force and the effective mass on the predictions for 
spherical shell closures in superheavy nuclei.

We have introduced two new RMF forces: NL--Z2 and NL--VT1, both employing
the standard nonlinear \emph{ansatz} for the Lagrangian, but NL-VT1
complemented with tensor couplings of the isoscalar and isovector
vector fields. Both are fitted to the same set of experimental data
as the recent Skyrme parameterizations SkI$x$. The tensor coupling
changes the relative distances between the single-particle states,
but it has no visible influence on spin-orbit splittings in heavy and
superheavy nuclei.

To test the predictive power of the models, we have compared the
experimental and calculated single-particle spectra in ${}^{208}{\rm Pb}$,
the heaviest known spherical doubly-magic nucleus so far.
Already in this nucleus used in the fit of all forces investigated here
we see large differences between calculations and experiment and
among the forces. States with large angular momentum are shifted to too 
small single-particle energies and none of the self-consistent models
gives the proper level ordering.

The predictions for shell closures are found to be sensitive to the 
isospin dependence of the spin-orbit interaction and the isoscalar 
effective mass. The uncertainties of these quantities in the description
of smaller nuclei amplify when going to large mass numbers, making
predictions for superheavy nuclei a demanding task.

The occurrence of proton shell closures in self-consistent models depends 
strongly on the neutron number (and vice versa), even when looking at 
spherical nuclei only. This effect can be explained in terms of 
single-particle spectra as a coupling of the spin-orbit field to the 
profile of the density distribution (of protons and neutrons separately) 
which undergoes dramatic changes in superheavy nuclei.
This is an effect of self-consistency, it cannot occur in models where the
density distribution has only a restricted degree of freedom like
the semi-microscopic ETFSI approach or has even no degree-of-freedom
at all like in case of macroscopic-microscopic models. 
In the region around ${}^{292}_{172}120$ all forces with small effective
mass predict a deep central depression of the nuclear density, which
induces an unusual shape of the spin-orbit potential that causes an 
additional state-dependence of the spin-orbit splitting. In some cases 
the usual level ordering of spin-orbit coupled states is even reverted. 

The change of the single-particle spectra of both protons and neutrons
when varying proton and neutron number is much larger in all 
self-consistent models than in non-self-consistent approaches, which was 
shown on the example of the Folded-Yukawa model.

The only self-consistent force which predicts \mbox{$Z=114$} for the
next spherical magic proton number is the extended Skyrme force SkI4. 
Although SkI4 gives a very good description of the binding energies in known 
(deformed) superheavy nuclei \cite{BuervenSuper} and reproduces
the kink in the isotopic shifts of the mean-square radii in heavy lead 
nuclei, it overestimates the spin-orbit splittings of proton states in heavy 
nuclei by $60-80 \%$. This discrepancy between this very good description
of bulk properties and a rather poor description of details of the 
single-particle spectra is yet to be understood.
Since a possible proton shell closure at \mbox{$Z=114$}
is caused by a large spin-orbit splitting, the unique prediction of SkI4
is very questionable. On the other hand, all RMF forces, which are in
very good agreement with experimental data for spin-orbit splittings
throughout the chart of nuclei predict a magic \mbox{$Z=120$}.

In summary this gives a strong argument that the next magic
proton number is \mbox{$Z=120$}, coupled with a magic neutron number
\mbox{$N = 172$}, still a far way to go from the heaviest presently 
known nucleus ${}^{277}_{165}112$.
%
%
\acknowledgments
The authors would like to thank S.~Hofmann and G.~M\"unzenberg for many 
valuable discussions. M.~B.\ thanks for the warm hospitality at the Joint 
Institute for Heavy-Ion Research, where a part of this work was done.
This work was supported in parts by Bundesministerium f\"ur Bildung 
und Forschung (BMBF), Project No.\ 06 ER 808, by Gesellschaft f\"ur 
Schwerionenforschung (GSI), by Graduiertenkolleg Schwerionenphysik,
by the U.S.\ Department of Energy under Contract No.\ DE--FG02--97ER41019 
with the University of North Carolina and Contract No.\ DE--FG02--96ER40963 
with the University of Tennessee and by the NATO grant SA.5--2--05 
(CRG.971541). The Joint Institute for Heavy Ion Research has as member 
institutions the University of Tennessee, Vanderbilt University, and the 
Oak Ridge National Laboratory; it is supported by the members and by the 
Department of Energy through contract No.\ DE--FG05--87ER40361 with the 
University of Tennessee. 
%
%
\begin{appendix}
\section{Details of the Mean-Field Models}
\label{Sect:Parameters}
%
%
\subsection{The Skyrme Energy Functional}
\label{Subsect:SHF}
%
%
\widetext
\begin{table*}[t!]
\caption{
\label{tab:SHFpar}
Parameters of the Skyrme energy interactions. The $t_i$, $x_i$, $b_4$, $b_4'$
and $\alpha$ are the parameters of the Skyrme functional 
(\protect\ref{eq:SkyrmeFunctional}), $\hbar^2/2m$ is the constant in the
calculation of the kinetic energy (\protect\ref{eq:kineticEnergy}). 
}
\begin{tabular}{lccccccc}
Parameter  & SkM*   & SkP & SkI1 & SkI3 & SkI4 & SLy6  & SLy7 \\ \tableline
$t_0$ $[{\rm MeV} \; {\rm fm}^3]$ 
      & $-2645.0$ & $-2931.70$ & $-1913.619$ & $-1762.88$ & $-1855.827$ & 
$-2479.50$  & $-2480.80$  \\
$t_1$ $[{\rm MeV} \; {\rm fm}^5]$
      &   410.0   &  320.662   &   439.809   &  561.608   & 473.829     &  462.180 
   & 461.290     \\
$t_2$ $[{\rm MeV} \; {\rm fm}^5]$
      & $-135.0$  & $-337.41$  &  2697.594   & $-227.090$ & 1006.855    & 
$-448.610$  & $-433.930$  \\
$t_3$ $[{\rm MeV} \; {\rm fm}^{3+\alpha}]$ 
      & 15595.0   & 18708.97   & 10592.267   & 8106.2     & 9703.607    & 13673.0  
   & 13669.0     \\
$x_0$ &   0.09    & 0.29215    & $-0.954536$ & 0.3083     & 0.405082    & 0.825    
   & 0.848       \\
$x_1$ &   0.0     & 0.65318    & $-5.782388$ & $-1.1722$  & $-2.889148$ & $-0.465$ 
   & $-0.492$    \\
$x_2$ &   0.0     & $-0.53732$ & $1.287379$  & $-1.0907$  & $1.325150$  & $-1.0$   
   & $-1.0$      \\
$x_3$ &   0.0     & 0.18103    & $-1.561421$ & 1.2926     & $1.145203$  & 1.355    
   & 1.393       \\
$b_4$ $[{\rm MeV} \; {\rm fm}^4]$  
      &  65.0     &  50.0      & 62.130      & 94.254     & 183.097     & 61.0     
   & 62.5        \\
$b_4'$ $[{\rm MeV} \; {\rm fm}^4]$ 
      &  65.0     &  50.0      & 62.130      & 0.0        & $-180.351$  & 61.0     
   & 62.5        \\
$\alpha$  
      & $1/6$     & $1/6$      & 0.25        & 0.25       & 0.25        & $1/6$    
   & $1/6$       \\
$\hbar^2/2m$ $[{\rm MeV} \; {\rm fm}^2]$
      & 20.733983 & 0.733983 & 20.7525     & 20.7525    & 20.7525     & 
20.73552985 & 20.73552985
\end{tabular}
\end{table*}
\narrowtext
%
%
The Skyrme energy functionals are constructed to be effective interactions 
for nuclear mean-field calculations. For even-even nuclei, the Skyrme
energy functional used in this paper
\begin{equation}
{\cal E} 
  =     {\cal E}_{\rm kin} [\tau ]
      + {\cal E}_{\rm Sk}  [\rho, \tau, {\bf J}]
      + {\cal E}_{\rm C}   [\rho_p]
      - {\cal E}_{\rm cm}
\quad , 
\end{equation}
is composed of the functional of the kinetic energy ${\cal E}_{\rm kin}$,
the effective functional for the strong interaction ${\cal E}_{\rm Sk}$
and the Coulomb interaction ${\cal E}_{\rm C}$ including the exchange term 
in Slater approximation, and the correction for spurious center-of-mass 
motion ${\cal E}_{\rm cm}$. The energy functionals are the spatial
integrals of the corresponding Hamiltonian densities ${\cal H}$
\begin{equation}
{\cal E}[\rho, \tau, {\bf J}]
= \int \! {\rm d}^3 r \; 
  {\cal H} [\rho({\bf r}), \tau({\bf r}), {\bf J}({\bf r})]
\quad .
\end{equation}
The actual functionals are given by
\begin{eqnarray}
\label{eq:kineticEnergy}
{\cal H}_{\rm kin}
& = & \frac{\hbar^2}{2m} \tau
      \\
{\cal H}_{\rm C}
& = & \frac{e^2}{2} \int \! {\rm d}^3 r'
      \frac{\rho_{\rm p} (\textbf{r}) \rho_{\rm p} (\textbf{r}') }
           {|\textbf{r} - \textbf{r}'|}
      - \frac{3e^2}{4} \left( \frac{3}{\pi} \right)^{1/3} \! \!
        \rho_{\rm p}^{4/3}
      \\
\label{eq:SkyrmeFunctional}
{\cal H}_{\rm Sk} 
& = &       \frac{b_0}{2}  \rho^2
          + b_1            \rho \tau
          - \frac{b_2}{2}  \rho \Delta \rho 
          + \frac{b_3}{3}  \rho^{\alpha +2}
      \nonumber \\  
&   & - \sum_q \bigg[
            \frac{b'_0}{2} \rho_q^2
          + b'_1           \rho_q \tau_q 
          - \frac{b'_2}{2} \rho_q \Delta \rho_q     
          + \frac{b'_3}{3} \rho^\alpha \rho_q^2
      \bigg]
      \nonumber \\  
&   & + {\cal H}_{\rm LS}
\end{eqnarray}
with various possibilities for the spin-orbit interaction
\begin{eqnarray}
{\cal H}_{\rm LS}^{({\rm std})}
& = & - b_4 \Big(   \rho \nabla \cdot {\bf {J}}
                  + \sum_q \rho_q \nabla \cdot {\bf J}_q \Big)
      \quad , \\
{\cal H}_{\rm LS}^{({\bf J})}
& = & {\cal H}_{\rm LS}^{({\rm std})}
      + c_1 {\bf J}^2 - c_1'  \sum_q {\bf J}^2_q
      \quad , \\
{\cal H}_{\rm LS}^{({\rm ext})}
& = & - b_4  \; \rho \nabla \cdot {\bf {J}}
      - b'_4 \sum_q \rho_q \nabla \cdot {\bf J}_q
\quad .
\end{eqnarray}
${\cal H}_{\rm LS}^{({\rm std})}$ is reproduced from 
${\cal H}_{\rm LS}^{({\rm ext})}$ setting \mbox{$b'_4 = b_4$}.

%
%
\widetext
\begin{table*}[t!]
\caption{
\label{tab:RMFpar}
Parameters of the RMF forces used in this investigation. The mass of the
isovector vector-field ${m_\rho=763\ {\rm MeV}}$ is not fitted and is the 
same for all forces.}
\begin{tabular}{lddddddccc}
Force & $m_N$
      & $m_\sigma$
      & $m_\omega$
      & $g_\sigma$
      & $g_\omega$
      & $g_\rho$
      & $b_2$
      & $b_3$
      &         \\ 
& (MeV) & (MeV) & (MeV) & & & (fm$^{-1}$) & & \\ \tableline
NL3       & 939.0 & 508.194 & 782.501 & 10.2170  & 12.8680 & 4.47400 & $-10.4310$ 
& $-28.8850$  & \\
NL-Z      & 938.9 & 488.67  & 780.0   & 10.0553  & 12.9086 & 4.84944 & $-13.5072$ 
& $-40.2243$  & \\
NL-Z2     & 938.9 & 493.150 & 780.0   & 10.1369  & 12.9084 & 4.55627 & $-13.7561$ 
& $-41.4013$  & \\
NL-VT1    & 938.9 & 484.307 & 780.0   & 9.81307  & 12.6504 & 4.63432 & $-13.2808$ 
& $-38.0773$  & $f_\omega/g_\omega=-0.102703$ \\
& & & & & & & & & $f_\rho/g_\rho=-4.71143$ \\  
\end{tabular}
\end{table*}
\narrowtext
%
%

The local density $\rho_q$, kinetic density $\tau_q$
and spin-orbit current ${\bf J}_q$ entering the functional 
are given by
\begin{eqnarray}
\rho_q
& = & \sum_{k \in \Omega_q} v_k^2 \; | \psi_k |^2
      \quad , \nonumber \\
\tau_q
& = & \sum_{k \in \Omega_q} v_k^2 \; | \nabla \psi_k |^2
      \quad , \\
{\bf J}_q
& = & - {\textstyle \frac{i}{2}} \sum_{k \in \Omega_q} v_k^2 \;
      \Big[ \psi_k^\dagger \, \nabla \times \hat\sigma \, \psi_k
            - ( \nabla \times \hat\sigma \, \psi_k )^\dagger \psi_k
      \Big]
      \quad , \nonumber
\end{eqnarray}
with $q \in \{p,n\}$. Densities without index denote total densities,
e.g.\ \mbox{$\rho = \rho_{\rm p} + \rho_{\rm n}$}. The $\psi_k$
are the single-particle wavefunctions and $v_k^2$ the
occupation probabilities calculated taking the residual
pairing interaction into account, see Appendix~\ref{Subsect:PairFunc}.
The parameters $b_i$ and $b'_i$ used in the above definition
are chosen to give a most compact formulation of the energy functional,
the corresponding mean-field Hamiltonian and residual interaction
\cite{Rei92b}. They are related to the more commonly used Skyrme force 
parameters $t_i$ and $x_i$ by
\begin{eqnarray}
   b_0 & = &   t_0 \big( 1 + {\textstyle \frac{1}{2}} x_0 \big) \quad , 
               \nonumber \\
   b'_0& = &   t_0 \big( {\textstyle \frac{1}{2}} + x_0 \big)   \quad ,  
               \nonumber \\
   b_1 & = &  {\textstyle \frac{1}{4}} 
              \big[  t_1 \big( 1 + {\textstyle \frac{1}{2}} x_1 \big)
                    +t_2 \big( 1 + {\textstyle \frac{1}{2}} x_2 \big)
              \big] \quad , \nonumber \\
   b'_1& = &  {\textstyle \frac{1}{4}} 
              \big[  t_1 \big({\textstyle \frac{1}{2}} + x_1 \big)
                    -t_2 \big({\textstyle \frac{1}{2}} + x_2 \big)
              \big] \quad , \nonumber \\
   b_2 & = &  {\textstyle \frac{1}{8}} 
              \big[ 3t_1 \big( 1 + {\textstyle \frac{1}{2}} x_1 \big)
                    -t_2 \big( 1 + {\textstyle \frac{1}{2}} x_2 \big)
              \big] \quad , \nonumber \\
   b'_2& = &  {\textstyle \frac{1}{8}} 
              \big[ 3t_1 \big( {\textstyle \frac{1}{2}} + x_1 \big)
                    +t_2 \big( {\textstyle \frac{1}{2}} + x_2 \big)
              \big] \quad , \nonumber \\
   b_3 & = &  {\textstyle \frac{1}{4}} 
              t_3 \big( 1 + {\textstyle \frac{1}{2}} x_3 \big) \quad ,
              \nonumber \\
   b'_3& = &  {\textstyle \frac{1}{4}} 
              t_3 \big( {\textstyle \frac{1}{2}} + x_3 \big)   \quad ,
              \nonumber \\
   c_1 & = &  -{\textstyle \frac{1}{8}} 
              ( t_1 x_1 + t_2 x_2 )
              \quad , \nonumber \\
   c'_1& = &  -{\textstyle \frac{1}{8}} 
              (  t_1 - t_2 )
              \quad , 
\end{eqnarray}
The actual parameters for the parameterizations used in this paper
are summarized in Table~\ref{tab:SHFpar}.

The single-particle Hamiltonian is derived variationally from the
energy functional. One obtains
\begin{equation}
\label{eq:SHF:hamiltonian}
\hat{h}_q
= - \nabla \cdot B_q \nabla + U_q - i {\bf W}_q \cdot \nabla \times \hat\sigma
\end{equation}
with the mean fields
\begin{equation}
B_q
= \frac{\delta {\cal E}}{\delta \tau_q}
  \quad , \quad 
U_q 
= \frac{\delta {\cal E}}{\delta \rho_q}
  \quad , \quad
{\bf W}_q 
= \frac{\delta {\cal E}}{\delta {\bf J}_q}
  \quad .
\end{equation}
For all forces, a center-of-mass correction is employed. For the 
SkI$x$ and SLy$x$ forces it is calculated perturbatively by subtracting
\begin{equation}
\label{eq:Ecm}
{\cal E}_{\rm cm}
= \frac{1}{2mA} 
   \langle\hat\textbf{P}\rule{0pt}{6.8pt}^2_{\rm cm} \rangle
\end{equation}
from the Skyrme functional after the convergence of the Hartree-Fock 
equations, while for SkM* and SkP only the diagonal direct terms 
in (\ref{eq:Ecm}) are considered self-consistently in the variational
equation \cite{SkyrmeFit}. For all but the SLy$x$ forces this
is the procedure used in the original fit. For SLy6 and SLy7 the
microscopic correction (\ref{eq:Ecm}) was considered in the variational
equations and therefore giving a contribution to the single-particle
energy. However, for large nuclei as discussed here the contribution 
of (\ref{eq:Ecm}) to the single-particle energies is negligible 
because the matrix elements are weighted with $1/(2mA)$ compared 
to the contributions from the energy functional.
We have therefore omitted this feature and follow the suggestion
of \cite{SLyx} to use the perturbatively calculated correction from 
(\ref{eq:Ecm}) instead.
%
%
\subsection{Relativistic Mean-Field Model}
\label{Subsect:RMF}
For the sake of a covariant notation, it is better to provide the
basic functional in the relativistic mean-field model as an
effective Lagrangian ${\cal L}$. For the present version of the RMF 
used in this study, we can summarize it as
\begin{equation}
{\cal L}_{\rm RMF} 
= {\cal L}_{\rm N} + {\cal L}_{\rm M} + {\cal L}_{\rm NM} 
      + {\cal L}_{\rm nonl} + {\cal L}_{\rm em}
\quad ,
\end{equation}
where ${\cal L}_{\rm N}$ is the free Dirac Lagrangian for the nucleons with
nucleon mass $m_{\rm N}$, equally for protons and neutrons
\begin{equation}
{\cal L}_{\rm N} 
= \sum_{k\in\Omega} v_k^2 \;
      \bar\psi_k
      \left( i \gamma_\mu \partial^\mu - m_{\rm N}
      \right) \psi_k
\quad .
\end{equation}
The Lagrangians of the fields and their couplings to the nucleons 
are given by
\begin{eqnarray}
{\cal L}_{\rm M} 
& = & {\textstyle\frac{1}{2}}
      (\partial_\mu \Phi_\sigma \partial^\mu \Phi_\sigma 
      - m_\sigma^2 \Phi^2)
      \nonumber \\ 
&   & 
      - {\textstyle\frac{1}{2}}
        \Big[ {\textstyle\frac{1}{2}}
              (\partial_\mu\Phi^{\mbox{}}_{\omega,\nu}
               -\partial_\nu\Phi^{\mbox{}}_{\omega,\mu}) \;
              \partial^\mu\Phi_\omega^{\nu}
              - m_\omega^2 \Phi_{\omega,\mu}^{\mbox{}} \Phi_\omega^\mu
        \Big]
       \nonumber \\ 
&   & 
      - {\textstyle\frac{1}{2}} 
        \Big[ {\textstyle\frac{1}{2}}
              (\partial_\mu\vec{\Phi}^{\mbox{}}_{\rho,\nu}
               -\partial_\nu\vec{\Phi}^{\mbox{}}_{\rho,\mu})\cdot
               \partial^\mu\vec{\Phi}_\rho^{\nu}
              - m_\omega^2 \vec{\Phi}_{\rho,\mu}^{\mbox{}}\cdot
             \vec{\Phi}_\rho^\mu
        \Big]
        \nonumber\\ 
{\cal L}_{\rm NM} 
& = & - g_\sigma \Phi_\sigma \rho_{\rm s}
      - g_\omega \Phi_{\omega,\mu} \rho^{\mu} 
      - g_\rho \vec{\Phi}_{\rho,\mu}\cdot\vec{\rho}^{\,\mu}, 
      \nonumber \\ 
{\cal L}_{\rm nonl}
& = & {\cal U}_{\sigma\sigma}[\Phi_\sigma],
      \nonumber  \\
{\cal L}_{\rm em} 
& = & - {\textstyle\frac{1}{4}} 
      F_{\mu\nu}F^{\mu\nu} 
      - e A_\mu \rho_p^\mu,
\end{eqnarray}
The model includes couplings of the scalar-isoscalar ($\Phi_\sigma$),
vector-isoscalar ($\Phi_{\omega, \mu}$), vector-isovector 
($\vec{\Phi}_{\rho, \mu}$), and electro-magnetic ($A_\mu$) field to the
corresponding scalar-isoscalar ($\rho_{\rm s}$), vector-isoscalar 
($\rho^\mu$) and vector-isovector ($\vec\rho^\mu$) densities of the nucleons
as well as the proton density $\rho_p^\mu$, which are defined as
\begin{eqnarray}
\rho_{\rm s} 
& = & \sum_{k\in\Omega} v_k^2 \; \bar\psi_k\psi_k,
      \nonumber \\
\rho^\mu 
& = & \sum_{k\in\Omega} v_k^2 \; \bar\psi_k\gamma_\mu\psi_k,
      \nonumber \\
\vec\rho^{\,\mu} 
& = & \sum_{k\in\Omega} v_k^2 \; \bar\psi_k\vec\tau\gamma_\mu\psi_k,
      \nonumber \\
\rho_p^\mu 
& = & \sum_{k\in\Omega_p} v_k^2 \; \bar\psi_k\gamma_\mu\psi_k .
\end{eqnarray}
${\cal U}_{\sigma \sigma}$ is the nonlinear self\/interaction of the
scalar-isoscalar field. All forces used in this paper employ
the standard {\em ansatz} \cite{Rei89}
\begin{equation}
{\cal U}_{\sigma\sigma}
 =  -{\textstyle \frac{1}{3}} b_3 \Phi_\sigma^3
      - {\textstyle \frac{1}{4}} b_4 \Phi_\sigma^4.
\end{equation}
In case of the parameterset NL-VT1 also a tensor coupling between the
nucleons and the vector fields is considered, which can be
written as
\begin{equation}
{\cal L}_{\rm NM}^{\rm t}
=   \frac{f_\omega}{2m_{\rm N}} \Phi_{\omega, \mu} \rho_{\rm t}^\mu
  + \frac{f_\rho}{2m_{\rm N}} \vec\Phi_{\rho, \mu} \cdot 
    \vec\rho_{\rm t}^{\,\mu}
\end{equation}
with the densities
\begin{eqnarray}
\rho_{\rm t}^\mu
& = & \partial_\nu \sum_{k \in \Omega} v_k^2 \bar\psi_k
      \sigma^{\mu \nu} \psi_k, \nonumber \\
\vec\rho_{\rm t}^{\,\mu}
& = & \partial_\nu \sum_{k \in \Omega} v_k^2 \bar\psi_k 
      \sigma^{\mu \nu} \vec\tau \psi_k ,
\end{eqnarray}
where \mbox{$\sigma^{\mu \nu} = (i/2) [ \gamma^\mu, \gamma^\nu]$}.
The masses $m_i$ and coupling constants of the fields are the free
parameters of the RMF which have to be adjusted to experimental data.
The actual parameters of the parameterizations used here are given in
Table~\ref{tab:RMFpar}. The equation of motion of the single-particle
states is derived from a variational principle
\begin{equation}
\epsilon_k \gamma_0 \psi_k
= \Big[ - i {\boldmath \gamma} \cdot \nabla 
        + m_{\rm N}
        + {\cal S}
        + \gamma_\mu {\cal V}^\mu 
  \Big] \psi_k
\end{equation}
where \mbox{${\cal S} = g_\sigma \Phi_\sigma$} and 
${\cal V}_\mu = g_\omega \Phi_{\omega, \mu} 
                      + {\textstyle\frac{1}{2}} g_\rho 
                        \vec{\Phi}_{\rho, \mu} \cdot \vec{\tau}
                      + {\textstyle\frac{1}{2}} e A_\mu ( 1 + \tau_0) 
$ are the scalar and vector field respectively. A more detailed
description of the model can be found in \cite{Rei89}.

For the residual pairing interaction and the center-of-mass correction the 
same non-relativistic approximation is used as in the SHF model, for NL-Z, 
NL-Z2 and NL-VT1 by subtracting perturbatively the full microscopic 
correction (\ref{eq:Ecm}), while for NL3 the harmonic oscillator estimate 
\mbox{$E_{\rm c.m.}=\frac{3}{4} \; 41 A^{-1/3} \; {\rm MeV}$} is
subtracted as done in the fit of these parameter sets.
%
%
\subsection{Pairing Energy Functional}
\label{Subsect:PairFunc}
%
%
\begin{table}[b!]
\caption{Pairing strength $V_{\rm n}$ for the neutrons
and $V_{\rm p}$ for the protons for the mean--field forces
used in this study. $m^\ast/m$ is the isoscalar effective
mass in infinite nuclear matter. Note that the absolute value
of the pairing strength decreases with increasing effective mass.
}
\label{tableVpair}
\begin{tabular}{ldll}
Force & $m^\ast/m$ & $V_{\rm n}$ [MeV fm$^3$] & $V_{\rm p}$ [MeV fm$^3$] 
\\ \tableline
SkM*     & 0.789   & $-276$   & $-292$ \\
SkP      & 1.0     & $-241$   & $-265$ \\
SkI1     & 0.693   & $-320$   & $-305$ \\  
SkI3     & 0.574   & $-340$   & $-351$ \\
SkI4     & 0.650   & $-310$   & $-324$ \\
SLy6     & 0.689   & $-308$   & $-320$ \\ 
SLy7     & 0.688   & $-308$   & $-320$ \\ \tableline
NL3      & 0.595   & $-329$   & $-342$ \\
NL-Z     & 0.583   & $-349$   & $-351$ \\
NL-Z2    & 0.583   & $-343$   & $-350$ \\
NL-VT1   & 0.600   & $-340$   & $-346$       
\end{tabular}
\end{table}
%
%
Pairing is treated in BCS approximation using a delta pairing force
\cite{Ton79,Krieger}, leading to the pairing energy functional
\begin{equation}
{\cal E}_{\rm pair}
= \frac{1}{4} \sum_{q = \{ {\rm p, n} \} } V_q \int \! {\rm d}^3 r \; \chi_q^2
\quad ,
\end{equation}
where \mbox{$\chi_q = -2 \sum_{k \in \Omega_q>0} f_k u_k v_k \; | \psi_k |^2$}
is the pairing density including state-dependent cut-off factors
$f_k$ to restrict the pairing interaction to the vicinity of the
Fermi surface \cite{pairStrength}. $v_k^2$ is the occupation probability 
of the given single-particle state and $u_k^2 = 1 - v_k^2$. 
The strengths $V_{\rm p}$ for protons and $V_{\rm n}$ for neutrons depend on 
the actual mean-field parameterization. They are optimized by fitting for
each parameterization separately the pairing gaps in isotopic and
isotonic chains of semi-magic nuclei throughout the chart of
nuclei. The actual values can be found in Table~\ref{tableVpair}.
The pairing-active space $\Omega_q$ is chosen to embrace one additional shell
of oscillator states above the Fermi energy with a smooth cutoff weight, 
see \cite{pairStrength} for details.
%
%
\subsection{The Folded-Yukawa Single-Particle Potential}
\label{Subsect:FY}
We present here only the details needed for our discussion. A more
detailed discussion of the parameterization of the potentials can be 
found in \cite{FRDM} and references therein.
The single-particle Hamiltonian of the Folded-Yukawa single-particle
model has the same structure as the one of the Skyrme-Hartree-Fock model
 (\ref{eq:SHF:hamiltonian}), but instead of calculating the potentials 
self-consistently from the actual density distributions, a parameterized 
guess for the functional form of the potentials is used. The nucleons 
have an effective mass of \mbox{$m^*/m = 1$} without any radial dependence, 
therefore $B$ is simply given by \mbox{$B = \hbar^2/2m$}. The 
single-particle potential $U$ is calculated from the folding of a 
Yukawa function with the sharp nuclear surface
\begin{equation}
U_q ({\bf r})
= - \frac{V_0}{4 \pi a^3} \int_V \! {\rm d}^3 r' 
  \frac{e^{-|{\bf r}-{\bf r}'|/a}}{|{\bf r}-{\bf r}'|/a}
\end{equation}
where the integration is performed oder the nuclear volume. Finally, 
the spin-orbit potential is given by the derivative of the
nuclear potential
\begin{equation}
{\bf W}_q  ({\bf r})
= - \lambda_q (A) \left( \frac{\hbar}{2m} \right)^2 
    \nabla \, U_q 
\end{equation}
with the coupling constants $\lambda_{\rm p} = 28.0 + 6.0 \; A / 240$ 
and $\lambda_{\rm p} = 31.5 + 4.5 \; A / 240$.
\end{appendix}
%
%

%
%


\newpage

\begin{references}

\bibitem{Mye66a}
  W. D. Myers, W. J. Swiatecki,
  Nucl. Phys. \textbf{81}, 1 (1966).

\bibitem{SuperNils}
   S. G. Nilsson, C. F. Tsang, A. Sobiczewski, Z. Szymanski, 
   S. Wycech, C. Gustafson, I.--L. Lamm, P. M\"oller, and B. Nilsson,
   Nucl. Phys. \textbf{A131}, 1 (1969).

\bibitem{Mosel} 
   U. Mosel and W. Greiner, 
   Z. Phys. \textbf{222}, 261 (1969).

\bibitem{Fiz72a}
  E. O. Fizet, J. R. Nix,
  Nucl. Phys. \textbf{A193}, 647 (1972).

\bibitem{Bra72a}
  M. Brack, J. Damg{\aa}rd, A. S. Jensen, H. C. Pauli, V. M. Strutinsky, 
  and C. Y. Wong, 
  Rev. Mod. Phys. \textbf{44}, 320 (1972).

\bibitem{Z111}
   S. Hofmann, V. Ninov, F. P. Hessberger,  P. Armbruster, 
   H. Folger, G. M\"unzenberg, H. J. Sch\"ott,  A. G. Popeko, 
   A. V. Yeremin, A. N. Andreyev, S. Saro, R. Janik, and M. Leino, 
   Z. Phys. \textbf{A350}, 277 (1995) and 
   Z. Phys. \textbf{A350}, 281 (1995).

\bibitem{Berkley}
  A. Ghiorso, D. Lee, L. P. Somerville, W. Loveland, J. M. Nitschke, 
  W. Ghiorso, G. T. Seaborg, P. Wilmarth, R. Leres, A. Wydler, M. Nurmia, 
  K. Gregorich, K. Czerwinski,  R. Gaylord, T. Hamilton,  N. J. Hannink, 
  D. C. Hoffman, C. Jarzynski, C. Kacher, B. Kadkhodayan, S. Kreek, 
  M. Lane, A. Lyon, M. A. McMahan, M. Neu, T. Sikkeland, W. J. Swiatecki, 
  A. T\"urler, J. T. Walton, and S. Yashita, 
  Nucl. Phys. \textbf{A583}, 861c  (1995); 
  Phys. Rev. {\bf C51}, R2293 (1995).

\bibitem{Z112}
   S. Hofmann, V. Ninov, F. P. Hessberger,  P. Armbruster, 
   H. Folger, G. M\"unzenberg, H. J. Sch\"ott,  A. G. Popeko, 
   A. V. Yeremin, S. Saro, R. Janik, and M. Leino, 
   Z. Phys. \textbf{A354}, 229 (1996).

\bibitem{Dubna2}
   Yu. A. Lazarev, Yu. V. Lobanov, Yu. Ts. Oganessian, V. K. Utyonkov,
   F. Sh. Abdullin, A. N. Polyakov, J. Rigol, I. V. Shirokovsky,
   Yu. S. Tsyganov, S. Iliev, V. G. Subbotin, A. M. Sukhov,
   G. V. Buklanov, B. N. Gikal, V. B. Kutner, A. N. Mezentsev,
   K. Subotic, J. F. Wild, R. W. Lougheed, and K. J. Moody,
   Phys. Rev. C \textbf{54}, 620 (1996).

\bibitem{SHReview}
  S. Hofmann, 
  Rep. Prog. Phys. \textbf{61}, 639 (1998).

\bibitem{Sob87}
  A. Sobiczewski, Z. Patyk, and S. {\'C}wiok, 
  Phys. Lett. \textbf{186}, 6 (1987).

\bibitem{Sob89}
  A. Sobiczewski, Z. Patyk, and S. {\'C}wiok, 
  Phys. Lett. \textbf{224}, 1 (1989).

\bibitem{Pat89a}
  Z. Patyk, A. Sobiczewski, P. Armbruster, and K.--H. Schmidt, 
  Nucl. Phys. \textbf{A491}, 267 (1989).

\bibitem{Patyk} 
   Z. Patyk and A. Sobiczewski, 
   Nucl. Phys. \textbf{A533}, 132 (1991).

\bibitem{Mol92a}
   P. M\"oller and J. R. Nix, 
   Nucl. Phys. \textbf{A549}, 84 (1992).

\bibitem{Mol94a}
   P. M\"oller and J. R. Nix,
   J. Phys. \textbf{G20}, 1681 (1994).

\bibitem{Laz94}
  Yu. A. Lazarev, Yu. V. Lobanov, Yu. S. Oganessian, V. K. Utyonkov, 
  F. Sh. Abdulin, G. V. Buklanov, B. N. Gikal, S. Iliev,  A. N. Mezentsev, 
  V. N. Polyakov, I. M. Sedykh, I. V. Shirokovsky, V. G. Subbotin, 
  A. M. Sukhov, Yu. S. Tsyganov, V. E. Zhuchko, R. W. Lougheed, 
  K. J. Moody, J. F. Wild, E. K. Hulet, and J. H. McQuaid,
  Phys. Rev. Lett. \textbf{73}, 624 (1994).

\bibitem{Reiter}
  P. Reiter, T. L. Khoo, C. J. Lister, D. Seweryniak, I. Ahmad, 
  M. Alcorta, M. P. Carpenter, J. A. Cizewski, C. N. Davids,
  G. Gervais, J. P. Greene, W. F. Henning, R. V. F. Janssens, 
  T. Lauritsen, S. Siem, A. A. Sonzogni, D. Sullivan, J. Uusitalo,
  I. Wiedenhöver, N. Amzal, P. A. Butler, A. J. Chewter, K. Y. Ding,
  N. Fotiades, J. D. Fox, P. T. Greenlees, R.-D. Herzberg, G. D. Jones, 
  W. Korten, M. Leino, and K. Vetter,
  Phys. Rev. Lett. \textbf{82}, 509 (1999).

\bibitem{Cwi83}
  S. {\'C}wiok, V. V. Pashkevich, J. Dudek, and W. Nazarewicz, 
  Nucl. Phys. \textbf{A410}, 254 (1983).

\bibitem{Cwi85}
  S. {\'C}wiok, Z. {\L}ojewski, and V. V. Pashkevich,  
  Nucl. Phys. \textbf{A444}, 1 (1985).

\bibitem{Bon86a}
  K. B\"oning, Z. Patyk, A. Sobiczewski, and S. \'Cwiok, 
  Z. Phys. \textbf{A325}, 479 (1986).

\bibitem{Smo95a}
  R. Smola{\'n}cuk, J. Skalski, und A. Sobiczewski, 
  Phys. Rev. C \textbf{52}, 1871 (1995).

\bibitem{Vau70b}
  D. Vautherin, M. V\'en\'eroni and D. M. Brink, 
  Phys. Lett. \textbf{B33}, 381 (1970).

\bibitem{Bei74a}
  M. Beiner, H. Flocard, M. V\'en\'eroni, and P. Quentin, 
  Proceedings of the 27th Nobel Symposium, 
  Super Heavy Elements --
     Theoretical Predictions and Experimental Generation, 
  Ronneby, Sweden (S. G. Nilsson und N. R. Nilsson, eds.), 
  Physica Scripta \textbf{10A}, 84 (1974).

\bibitem{Ton80}
  F. Tondeur,  
  Z. Phys. \textbf{A297}, 61 (1980).

\bibitem{Gam90}
  Y. K. Gambhir, P. Ring, and A. Thimet, 
  Ann. Phys. (N.Y.) \textbf{198}, 132 (1990).

\bibitem{Boe93}
  H. F. Boersma, 
  Phys. Rev. \textbf{C48}, 472 (1993).

\bibitem{Berger}
   J.--F. Berger, L. Bitaud, J. Decharg\'e, M. Girod, 
   and S. Peru--Dessenfants,
   Proceedings of the International Workshop XXXIV on
   Gross Properties of Nuclei and Nuclear Exitations,
   Hirschegg, Austria, January 1996.
   (H. Feldmeier, J. Knoll, and W. N\"orenberg, edts.)  
   Gesellschaft f\"ur SChwerionenforschung, Darmstadt, 1996, p 56.

\bibitem{Ring}
   G. A. Lalazissis, M. M. Sharma, P. Ring, and Y. K. Gambir,
   Nucl. Phys. \textbf{A608}, 202 (1996).

\bibitem{Naz}
   S. \'Cwiok, J. Dobaczewski, P.--H. Heenen, P. Magierski, 
   and W. Nazarewicz,
   Nucl. Phys. \textbf{A611}, 211 (1996).

\bibitem{RutzSuper}
   K. Rutz, M. Bender, T. B\"urvenich, T. Schilling,
   P.--G. Reinhard, J. A. Maruhn, and W. Greiner,
   Phys. Rev. C \textbf{56}, 238 (1997).
                                              
\bibitem{BuervenSuper}
   T. B\"urvenich, K. Rutz, M. Bender, P.--G. Reinhard, J. A. Maruhn, 
   and W. Greiner,
   EPJ \textbf{A 3}, 139 (1998).  

\bibitem{SHpes}
   M. Bender, K. Rutz, P.--G. Reinhard, J. A. Maruhn, and W. Greiner,
   Phys. Rev. C \textbf{58}, 2126 (1998).

\bibitem{refSHF}
   P. Quentin and H. Flocard, 
   Ann. Rev. Nucl. Part. Sci. \textbf{28}, 523 (1978).

\bibitem{WalSer}
   B. D. Serot and J. D. Walecka, 
   Adv. Nucl. Phys. \textbf{16}, 1 (1986).

\bibitem{Rei89}
   P.--G. Reinhard, 
   Rep. Prog. Phys. \textbf{52}, 439 (1989).

\bibitem{Rin95}
   P. Ring, 
   Prog. Part. Nucl. Phys. \textbf{37}, 193 (1996).

\bibitem{Gogny}
   J. Decharg\'e and D. Gogny,
   Phys. Rev. \textbf{21}, 1568 (1980).

\bibitem{HohKohn}
  P. Hohenberg and W. Kohn,
  Phys. Rev. \textbf{B136}, 864 (1964).

\bibitem{KohnSham}
  W. Kohn, L. J. Sham,
  Phys. Rev. A \textbf{140}, 1133 (1965).

\bibitem{ReiCorr}
  P.--G. Reinhard and C. Toepffer,
  Int. J. of Mod. Phys. \textbf{E3}, 435 (1994).

\bibitem{SkM*}
   J. Bartel, P. Quentin, M. Brack, C. Guet, and H.--B. H{\aa}kansson,
   Nucl. Phys. \textbf{A386}, 79 (1982).

\bibitem{SkP}
   J. Dobaczewski, H. Flocard, and J. Treiner, 
   Nucl. Phys. \textbf{A422}, 103 (1984).

\bibitem{Chabanat}
   E. Chabanat, 
   {\sl Interactions effectives pour des conditions extr\^emes d'isospin}, 
   Universit\'e Claude Bernard Lyon-1, Thesis 1995, LYCEN~T~9501, unpublished.

\bibitem{SLyx}
   E. Chabanat, P. Bonche, P. Haensel, J. Meyer, and R. Schaeffer,
   Nucl. Phys. \textbf{A635}, 231 (1998).

\bibitem{SkIx}
   P.--G. Reinhard and H. Flocard,  
   Nucl. Phys. \textbf{A584}, 467 (1995).

\bibitem{NL3}
   G. A. Lalazissis, J. K{\"o}nig, and P. Ring,
   Phys. Rev. C \textbf{55}, 540 (1997).

\bibitem{NLZ}
   M. Rufa, P.--G. Reinhard, J. A. Maruhn, W. Greiner, and M. R. Strayer,
   Phys. Rev. C \textbf{38}, 390 (1989).

\bibitem{SkyrmeFit}
   J. Friedrich and P.--G. Reinhard, 
   Phys. Rev. C \textbf{33}, 335 (1986).

\bibitem{Blaizot}
 J.~P. Blaizot, Phys. Rep. \textbf{64}, 171 (1980).

\bibitem{BGHrev}
 M. Brack, C. Guet, H.-B. Hakansson, Phys.Rep. \textbf{123}, 275 (1985).

\bibitem{dampgrad}
   V. Blum, G. Lauritsch, J. A. Maruhn, and P.--G. Reinhard,
   J. Comp. Phys. \textbf{100}, 364 (1992).

\bibitem{LS1}
 O. Haxel, J. H. D. Jensen and H. E. Suess,
 Phys. Rev. \textbf{75}, 1766 (1949).

\bibitem{LS2}
 M. Goeppert Mayer,
 Phys. Rev. \textbf{75}, 1969 (1949).

\bibitem{Nilssonbook}
   S. G. Nilsson and I. Ragnarsson, 
   \emph{Shapes and shells in nuclear structure},
   Cambridge University Press, Cambridge, 1995.

\bibitem{Duerr}
   H. P. Duerr,
   Phys. Rev. \textbf{103}, 469 (1956).

\bibitem{Bell}
   J. S. Bell and T. H. R. Skyrme,
   Philosophical Magazine \textbf{1}, 1055 (1965).
  
\bibitem{Sky59b}
  T. H. R. Skyrme,
  Nucl. Phys. \textbf{9}, 635 (1959).

\bibitem{Lal94a}
  G. A. Lalazissis, M. M. Sharma, J. K{\"o}nig, and P. Ring, 
  International Conference on ``Nuclear Shapes and Nuclear Structure at Low
  Excitation Energies'', Antibes (France) June 20--25, 1994 
  (Gif--sur--Yvette Cedex, France) 
  (M. Vergnes, D. Goutte, P.--H. Heenen, und J. Sauvage, eds.), 
  Editions Frontieres, 1994, p.~161.

\bibitem{Sha95a}
  M. M. Sharma, G. A. Lalazissis, J. K{\"o}nig, and P. Ring,  
  Phys. Rev. Lett. \textbf{74}, 3744  (1995).

\bibitem{FY}
  M. Bolsterli, E. O. Fiset, J. R. Nix, and J. L. Norton,
  Phys. Rev. C \textbf{5}, 1050 (1972).

\bibitem{WS}
  W. Nazarewicz, J. Dudek, R. Bengtsson, T. Bengtsson, and I. Ragnarsson,
  Nucl. Phys. \textbf{A 435}, 397 (1985).

\bibitem{Dob95}
  J. Dobaczewski, W. Nazarewicz, and T. R. Werner,
  Phys. Scr. \textbf{T56}, 15 (1995).

\bibitem{JamMah}
  M. Jaminon and C. Mahaux, 
  Phys. Rev. C {\bf 40}, 354 (1989).

\bibitem{oddNuclei}
  K. Rutz, M. Bender, P.--G. Reinhard, J. A. Maruhn, and W. Greiner,
  Nucl. Phys. \textbf{A634}, 67 (1998).

\bibitem{SPEexp}
  R. R. Kinsey {\em et al.}, 
  \emph{The NUDAT/PCNUDAT Program for Nuclear Data},
  9th International Symposium of Capture Gamma--Ray Spectroscopy and 
  Related Topics (Budapest, Hungary, October 1996). 
  Data extracted from the NUDAT database, version March 20, 1997, 
  National Nuclear Data Center WorldWideWeb site.

\bibitem{Hirschegg}
  M. Bender, T. B\"urvenich, K. Rutz, J. A. Maruhn, W. Greiner, and
  P.--G. Reinhard, 
  Proceedings of the International Workshop XXXVI on
  Gross Properties of Nuclei and Nuclear Exitations,
  Hirschegg, Austria, January 11--17 1998,
  (M. Buballa, W. N\"orenberg, J. Wambach, A. Wirzba, eds.)
  Gesellschaft f\"ur Schwerionenforschung, Darmstadt, 1998, p.~59.

\bibitem{Gatlinburg}
  M. Bender, K. Rutz, T. B\"urvenich, P.--G. Reinhard, J. A. Maruhn, 
  and W. Greiner,
  Proceedings of the Nuclear Structure '98 International Conference,
  Gatlinburg, Tennessee, August 10--15 1998.
  (to be published)

\bibitem{FRDM}
  P. M\"oller, J. R. Nix, W. D. Myers, and W. J. Swiatecki,
  At. Data Nucl. Data Tables \textbf{59}, 185 (1995).

\bibitem{Lew89}
  M. Lewitowicz, Yu. E. Penionzhkevich, A. G. Artukh, A. M. Kalinin, 
  V. V. Kamanin,
  S. M. Lukyanov, Nguyen Hoai Chau, A. C. Mueller, D. Guillemaud--Mueller, 
  R. Anne, D. Bazin, C. Detraz, D. Guerreau, M. G. Saint--Laurent, V. Borrel, 
  J. C. Jacmart, F. Pougheon, A. Richard, W. D. Schmidt--Ott, 
  Nucl. Phys. \textbf{A496}, 477 (1989).

\bibitem{Sor93}
  O. Sorlin, D. Guillemaud--Mueller, A. C. Mueller, V. Borrel, S. Dogny, 
  F. Pougheon, K.--L. Kratz, H. Gabelmann, B. Pfeiffer, A. Wohr, W. Ziegert, 
  Yu. E. Penionzhkevich, S. M. Lukyanov, V. S. Salamatin, R. Anne, C. Borcea, 
  L. K. Fifield, M. Lewitowicz, M. G. Saint--Laurent, D. Bazin, C. Detraz, 
  F.--K. Thielemann, and W. Hillebrandt, 
  Phys. Rev. C \textbf{47}, 2941 (1993).

\bibitem{Sch96}
  H. Scheit, T. Glasmacher, B. A. Brown, J. A. Brown, P. D. Cottle, 
  P. G. Hansen, R. Harkewicz, M. Hellstrom, R. W. Ibbotson, J. K. Jewell, 
  K. W. Kemper, D. J. Morrissey, M. Steiner, P. Thirolf, and M. Thoennessen, 
  Phys. Rev. Lett. \textbf{77}, 3967 (1996).

\bibitem{Gla97}
  T. Glasmacher, B. A. Brown, M. J. Chromik, P. D. Cottle, M. Fauerbach, 
  R. W. Ibbotson, K. W. Kemper, D. J. Morrissey, H. Scheit, D. W. Sklenicka, 
  and M. Steiner,
  Phys. Lett. B \textbf{395}, 163 (1997).

\bibitem{Wer96a}
  T. R. Werner, J. A. Sheikh, M. Misu, W. Nazarewicz, J. Rikovska, K. Heeger, 
  A. S. Umar, and M. R. Strayer,
  Nucl. Phys. \textbf{A597}, 327 (1996) and work cited therein.

\bibitem{Lal98a}
  G. A. Lalazissis, D. Vretanar, P. Ring, M. Stoitsov, and L. Robledo,
  preprint nucl-th/9807029.

\bibitem{Pea91a}
  J. M. Pearson, Y. Aboussir, A. K. Dutta, R. C. Nayak, M. Farine,
  and F. Tondeur,
  Nucl. Phys. \textbf{A528}, 1 (1991).

\bibitem{ETFSI}
  Y. Aboussir, J. M. Pearson, A. K. Dutta, and F. Tondeur,
  At. Data Nucl. Data Tables \textbf{61}, 127 (1995).

\bibitem{Rei92b}
  P.--G. Reinhard,
  Ann. Phys. (Leipzig) \textbf{1}, 632 (1992).

\bibitem{Ton79}
  F. Tondeur,
  Phys. Lett. \textbf{123B}, 139 (1983).

\bibitem{Krieger}
  S. J. Krieger, P. Bonche, H. Flocard, P. Quentin, and M. S. Weiss,
  Nucl. Phys. \textbf{A517}, 275 (1990).

\bibitem{pairStrength}
  M. Bender, P.--G. Reinhard, K. Rutz, and J. A. Maruhn,
  preprint, submitted to Phys. Rev. C (1998). 

\end{references}
\end{document}